\documentclass[english,aps,twocolumn,prd,superscriptaddress,showpacs,preprintnumbers, amsmath,amssymb,nofootinbib]{revtex4}

\usepackage{subfigure} 
\usepackage{amsmath,amssymb}
\usepackage{graphicx}
\usepackage{rotating}
\usepackage{bm}
\usepackage{dsfont}
\usepackage{slashed}
\usepackage{morefloats}
\usepackage{color}
\usepackage[dvips]{epsfig}     

\definecolor{refcol}{RGB}{0,0,150}
\usepackage{microtype}
\usepackage[colorlinks,linkcolor=refcol,citecolor=refcol,urlcolor=refcol]{hyperref}
\usepackage{breakurl}

\renewcommand{\sp}{\scriptscriptstyle}

\definecolor{comment}{rgb}{0.9,0,0}

\def\di{\displaystyle}

\def\bg{\begin{eqnarray}\begin{array}{rcl}\displaystyle}
\def\eg{\end{array} &\di    &\di   \end{eqnarray}}
\def\bm#1{\begin{eqnarray}\begin{array}{#1}\di}
\def\bmo#1{\begin{eqnarray*}\begin{array}{#1}\di}
\def\bml#1#2{\begin{eqnarray}\begin{array}{#1}\label{#2}\di}
\def\bgo{\begin{eqnarray*}\begin{array}{rcl}\displaystyle}
\def\ego{\end{array} &\di    &\di \nonumber  \end{eqnarray*}}

\def\btensor#1#2{\renew\left#1\begin{array}{#2}\di}
\def\brtensor#1#2#3{\ren#3\left#1\begin{array}{#2}}
\def\botensor#1#2{\renew\left#1\begin{array}{#2}}
\def\etensor#1{\end{array}\right#1}

\def\eq#1{(\ref{#1})}
\def\Eq#1{Eq.~(\ref{#1})}

\def\pa{\partial}

\def\d{\delta}
\def\det{{\rm det}}

\def\s0#1#2{\mbox{\small{$ \frac{#1}{#2} $}}}
\def\0#1#2{\frac{#1}{#2}}

\def\p{\partial\llap{/}}
\def\pa{\partial}

\def\e{\mbox{\boldmath$\epsilon$}}
\def\r{R^\psi}

\def\vf{\varphi}
\def\s{\sigma}	
\def\h{\eta}	
\def\D{\Delta}
\def\G{\Gamma}
\def\F{\Phi}
\def\J{\Psi}
\def\g{\gamma}
\def\p{\pi}
\def\k{\kappa}
\def\t{\tau}
\def\l{\lambda}

\def\r{\rho}
\def\m{\mu}
\def\ra{\rightarrow}

\def\f{{\mbox{\tiny $f$}}}

\def\fig#1{Fig.~\ref{#1}}

\newcommand{\bea}{\begin{eqnarray}}
\newcommand{\eea}{\end{eqnarray}}

\renewcommand{\Re}{{\rm Re}}

\def\ren#1{\renewcommand{\arraystretch}{#1}}

\def\renew{\renewcommand{\arraystretch}{1}}
\graphicspath{{./figures/}}

\definecolor{blue}{rgb}{0,0,1}

\definecolor{green}{rgb}{0,1,0}

\definecolor{red}{rgb}{1,0,0}

\renewcommand{\f}{\phi}

\newcommand{\tr}{\mathrm{tr}}

\newcommand{\be}{\begin{eqnarray}}
  \newcommand{\ee}{\end{eqnarray}}

\usepackage{babel}
\makeatother

\begin{document}

\title{The Phase Diagram of QC${}_2$D from Functional Methods}
%\date{\today}                                           
\author{Naseemuddin Khan}
\affiliation{Institut f\"ur Theoretische Physik, Heidelberg University, 
Philosophenweg 16, 62910 Heidelberg, Germany}
\affiliation{ExtreMe Matter Institute EMMI, GSI Helmholtzzentrum f\"ur 
Schwerionenforschung, Planckstr. 1, 64291 Darmstadt, Germany}
\author{Jan M.~Pawlowski}
\affiliation{Institut f\"ur Theoretische Physik, Heidelberg University, 
Philosophenweg 16, 62910 Heidelberg, Germany}
\affiliation{ExtreMe Matter Institute EMMI, GSI Helmholtzzentrum f\"ur 
Schwerionenforschung, Planckstr. 1, 64291 Darmstadt, Germany}
\author{Fabian Rennecke} 
\affiliation{Institut f\"ur Theoretische
  Physik, Justus-Liebig-Universit\"at Gie\ss en, Heinrich-Buff-Ring 16, 35392 Gie\ss
  en,Germany}
\author{Michael M.~Scherer}
\affiliation{Institut f\"ur Theoretische Physik, Heidelberg University, 
Philosophenweg 16, 62910 Heidelberg, Germany}

%%%%%%%%%%%%%%%%%%%%%%%%%%%%%%%%%%%
%%%%%%%%%%%%%%%%%%%%%%%%%%%%%%%%%%%

\begin{abstract}
  We study the phase diagram of two-color Quantum Chromodynamics at
  finite temperature and chemical potential. This is done within an
  effective low-energy description in terms of quarks, mesons and
  diquarks. Quantum, thermal and density fluctuations are taken into
  account with the functional renormalisation group approach. In
  particular, we establish the phenomenon of pre-condensation,
  affecting the location of the phase boundary to Bose-Einstein
  condensation. We also discuss the Silver Blaze property in the
  context of the functional renormalisation group.
\end{abstract}
\maketitle
%

%%%%%%%%%%%%%%%%%%%%%%%%%%%%%%%%%%%
%%%%%%%%%%%%%%%%%%%%%%%%%%%%%%%%%%%

\section{Introduction}
The investigation of the phase diagram of Quantum Chromodynamics (QCD)
at finite temperature and density is an area of very active
experimental and theoretical research~\cite{BraunMunzinger:2009zz}. So
far, theoretical investigations of the QCD phase structure have not
matured enough to provide quantitative predictions at finite
density. While ab initio lattice simulations at finite density are
hampered by the sign problem, ab intio continuum computations are
hampered by the so far missing access to all fluctuating degrees of
freedom at high density. Moreover, at high density we also expect
competing order effects that further increase the need for fully
quantitative computations.  In this situation the investigation of
QCD-like theories, that lack the above-mentioned problems, is an
interesting option for shedding light on particular aspects of finite
density QCD.  In the present work we study QCD with two colors,
$N_c=2$, and two quark flavors $N_f=2$. Two-color QCD (QC${}_2$D) has
no sign problem, and exhibits an additional meson-baryon symmetry
(Pauli-G\"ursey) not present in QCD with three colors. In QC${}_2$D
baryons are diquark states and the meson-baryon symmetry is related to
the bosonic nature of the baryons. Consequently QC${}_2$D allows for
both, the chirally broken mesonic phase of quark-antiquark pairs known
from QCD, and the (Bose-Einstein-) condensation of colorless
diquarks. In summary, despite its qualitative difference to QCD,
QC${}_2$D also shares many interesting similarities with
QCD. Moreover, its accessibility for lattice simulations makes it a
perfect testbed for continuum approaches to QCD, both for ab initio
computations and low-energy effective theories.

For the above reasons QC${}_2$D has recently attracted strong
interest. It has been studied within mean field theory and the chiral
Lagrangian
approach~\cite{Kogut:1999iv,Kogut:2000ek,Splittorff:2000mm,Splittorff:2001fy,
  Dunne:2002vb,Dunne:2003ji,Brauner:2006dv,Kanazawa:2009ks}, random
matrix models~\cite{Kanazawa:2009en} and within
Nambu--Jona-Lasinio--type
models~\cite{Kondratyuk:1991hf,Rapp:1997zu,Ratti:2004ra,Sun:2007fc,
  Brauner:2009gu,Andersen:2010vu,Harada:2010vy,Zhang:2010kn,Imai:2012hr,He:2010nb}. For
an even number of quark flavors $N_f$ the SU(2) gauge group provides a
positive path integral measure and thus avoids the occurrence of a
fermion-sign problem. This facilitates the investigation of the
QC${}_2$D phase diagram by lattice
simulations~\cite{Nakamura:1984uz,Hands:1999md,Hands:2000ei,Hands:2001ee,Muroya:2002jj,
  Chandrasekharan:2006tz,Hands:2006ve,Hands:2011ye,Cotter:2012mb,
  Hands:2012yy,Boz:2013rca,Scior:2015vra}.

In this work we employ the functional renormalisation group (FRG)
approach to the quark-meson-diquark model, \cite{Strodthoff:2011tz},
to study the phase diagram of two-color QCD. For QCD-related reviews
of the FRG see \cite{Litim:1998nf, Berges:2000ew, Polonyi:2001se,
  Pawlowski:2005xe, Gies:2006wv, Schaefer:2006sr, Rosten:2010vm,
  Braun:2011pp, vonSmekal:2012vx,Pawlowski:2014aha}. In the case of
QC${}_2$D a comprehensible FRG phase diagram for finite temperature
and density has previously been established by Strodthoff \emph{et
  al.} in Refs.~\cite{Strodthoff:2011tz,Strodthoff:2013cua}. There,
the flow equation for the effective potential was solved in the lowest
order of the derivative expansion. Such an approximation takes into
account momentum scale dependent multi-scattering processes of mesons
and baryons.

In the present work we significantly extend the truncation scheme in
\cite{Strodthoff:2011tz,Strodthoff:2013cua} by including
non-perturbative corrections to the classical dispersion relations of
quarks, mesons and diquarks by means of the respective running wave
function renormalisations. We also take into account the momentum
scale dependence of quark-meson and quark-diquark scattering processes
with running Yukawa couplings. It has been shown in QCD as well as in
the non-relativistic analogue of QC${}_2$D, namely the
non-relativistic BCS-BEC crossover, that these fluctuations play a
large quantitative r$\hat{\rm o}$le. For FRG work on this topic see
e.g.~\cite{Pawlowski:2014zaa, Diehl:2007ri, Diehl:2009ma,
  Floerchinger:2008qc, Floerchinger:2009pg, Scherer:2010sv,
  Boettcher:2012cm, Schnoerr:2013bk}, for reviews see
\cite{Scherer:2010sv,Boettcher:2012cm}. Here, we discuss their impact
on the phase structure of QC${}_2$D. This includes a careful
assessment of the Silver Blaze property. The latter is evaluated in
the context of renormalisation group flows. We also deduce the basic
requirements that approximations of $n$-point functions have to
fulfil in order to maintain the Silver Blaze property. Finally, we
also study pre-condensation in two-color QCD: This phenomenon can
appear when a physics system in the disorder phase is close to a
transition toward an ordered state.  Pre-condensation means that order
occurs at intermediate length scales or momentum scales. In the
present FRG setting this phenomenon is easily accessible due to the
successive integration of momentum modes. In summary we establish a
refined picture of the FRG phase diagram for QC${}_2$D.

The paper is organised as follows: In Sec.~\ref{sec:effmod}, we
introduce the quark-meson diquark (QMD) model as a low energy
effective theory for QC${}_2$D. In Sec.~\ref{sec:frg}, we review
basics of the FRG method and discuss the Silver Blaze property in the
present context, cf. Sec.~\ref{sec:SilvBl}. We further discuss the
truncation for the fluctuation analysis of the QMD-model, and derive
the flow equations for the effective potential, the wave function
renormalisations and the Yukawa couplings in Sec.~\ref{sec:setup}. The
results for the chiral and diquark condensation and related phase
diagrams are presented in Sec. \ref{sec:RES}, and are compared to the
literature. Technical details as well as an analysis of truncation
effects are given in the Appendix.

%%%%%%%%%%%%%%%%%%%%%%%%%%%%%%%%%%%
%%%%%%%%%%%%%%%%%%%%%%%%%%%%%%%%%%%

\section{Low-Energy description of QC${}_2$D}
\label{sec:effmod}

\vspace{-3.mm}
Here we outline the construction of our low-energy description of
QC${}_2$D with two quark flavors. By integrating out the gluons first,
we derive a low energy effective model in terms of quarks, mesons and
diquarks, \cite{Strodthoff:2011tz}, valid at momentum- or energy
scales $\Lambda \lesssim 1$ GeV. The quantum, thermal and density
fluctuations below $\Lambda$ are taken into account within the
functional renormalisation group approach.

\subsection{Quark-Meson-Diquark Model for QC${}_2$D}
\begin{figure}[t!]
 \center
 \includegraphics[width=0.9\columnwidth]{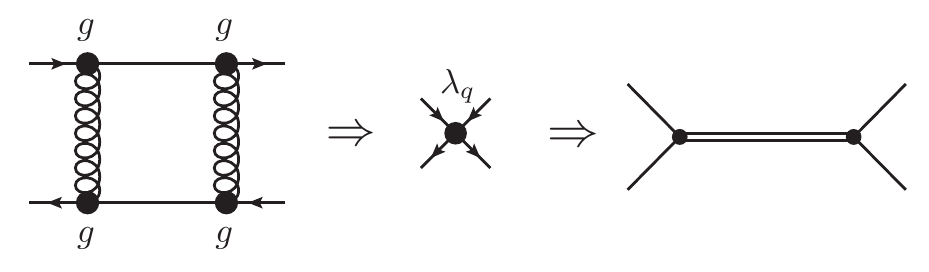}
 \caption{An effective four-fermi coupling (middle) is generated by
   gluon exchange (left). This process is the starting point to
   motivate effective models, where high energy degrees of freedom
   have been integrated out. By means of a Hubbard-Stratonovich
   transformation we introduce composite fields into our model as,
   e.g., represented by the double line in the diagram on the right.}
\label{fig:fourfermi} 
\end{figure}

The starting point is the microscopic QCD Lagrangian. Here we focus on
the matter part, 
\be \mathcal{L}_{\mathrm{mat}}= q^\dagger_L i\s_\m D_\m q_L+
q^\dagger_R i\s^\dagger_\m D_\m q_R+i m_q \bar q q
\,, \label{eq:chiralDirac} \ee
where the covariant derivative $D_\m=\pa_\m+igA_\m$ couples quarks to
gluons via the strong coupling $g$, where $A_\m=A_\m^at_a$ with the
gauge group generators $t_a$.  We define
$\s_\m=\left(i\s_j,\mathds{1}\right) $ with the Pauli matrices
$\s_j$. The $\gamma$-matrices are defined as
\begin{align}
  \gamma_j = \begin{pmatrix} 0& -i \s_j\\ 􏰉i \s_j& 0 \end{pmatrix},\;
  \gamma_0 = \begin{pmatrix} 0& \mathds{1}\\ \mathds{1} &
    0 \end{pmatrix},\; \gamma_5 = \begin{pmatrix} \mathds{1}& 0 \\ 􏰉0
    & -\mathds{1} \end{pmatrix}.
\end{align}
In the case of two colors, the gauge group generators (the Pauli
matrices) are pseudo-real, $t_a^*=t_a^T=-t_2t_at_2$. This property
leads to an extended $SU(2N_f)$ flavor symmetry, also known as
Pauli-G\"ursey symmetry. Naturally, various four-fermion interaction
channels are generated by gluon-exchange diagrams,
cf.~Fig.~\ref{fig:fourfermi}. In the present low energy effective
model we allow for interaction terms respecting the $SU(4)$ flavor
symmetry of QC${}_2$D with a scalar, a pseudo-scalar and a diquark
channel. Basically, this is a minimal two-flavor Nambu--Jona-Lasinio
(NJL) Lagrangian, supplemented with a diquark term,
\begin{eqnarray}\label{eq:NJLint}
  &\begin{aligned}
    \mathcal{L}_{\text{NJL}}&=\mathcal{L}^q_{\mathrm{kin}}+\l_q\left[(
      \bar{q}q)^2+(\bar{q} i\gamma_5\vec\tau q)^2\right. 
    \\ &\qquad\qquad
    \left. 
      +\big( \bar{q}i\gamma_5\tau_2 t_2C\bar{q}^T\big) \big( q^TCi
      \g_5 \t_2t_2 q\big) \right]\,,
\end{aligned}
\end{eqnarray}
where $C=\g_2\g_0$ is the charge conjugation operator and the $\t_i$'s
are Pauli matrices in flavor (isospin) space. The coupling constants
for the various channels are fixed by the $SU(4)$ flavor symmetry.
The low-energy properties of QC${}_2$D can be described most
efficiently by composite fields obtained with a Hubbard-Stratonovich
transformations. For the NJL model with the interaction Lagrangian in
\eq{eq:NJLint} this amounts to the introduction of a meson
field \mbox{$\f=(\vec{\pi},\sigma)$} which parametrises the pairing in
the the pseudo-scalar and the scalar channel as well as a complex
diquark field $\Delta$. A fluctuation analysis of the above action
leads to a propagation of the meson and diquark degrees of freedom,
represented by kinetic terms of mesons and diquarks as well as a
regeneration of four-fermi interactions. The latter can be rewritten
as corrections of the propagators and vertices of quarks and composite
fields leading to vanishing meson and diquark channels of the
four-fermi interaction at all cutoff scales. This dynamical
hadronisation,
\cite{Gies:2001nw,Pawlowski:2005xe,Floerchinger:2009uf}, has been
detailed in QCD in \cite{Braun:2014ata,Mitter:2014wpa}. In the
present work we neglect the related effects that is in particular
important for the high density part of the phase diagram, for a discussion se 
\cite{Pawlowski:2014aha}.

For the investigation of the theory at finite density a coupling to a
chemical potential $\m$ for the quarks as well as for the diquarks has
to be included. Putting everything together we are led to the
Lagrangian of the QMD-model,
\begin{align}\label{eq:QMD}
\begin{split}
  \mathcal{L}_{\text{QMD}} &= \bar{q}\left( i\slashed{\pa} +
    i\gamma_0\m +ih(
    \s+i\g_5\vec{\t}\cdot\vec{\p} )\right) q\\
  &\quad+\frac{1}{2}m^2\left(\vec{\p}^2+\s^2+|\D|^2\right)-c\s \\
  &\quad+\frac{h}{2}\left(i\D^* q^T C \gamma_5 \tau_2 t_2 q-{\rm h.c.}\right)\\
  &\quad+\frac{1}{2}(\partial_\mu\vec{\pi})^2 + \frac{1}{2}(\partial_\mu\sigma)^2+ \frac{1}{2}\left| \partial_i\D\right|^2\\
  &\quad+\frac{1}{2}
  \left(\partial_\tau-2\mu\right)\D^*\left(\partial_\tau+2\mu\right)\D\,.
\end{split}
\end{align}
The four-fermi interaction in \eq{eq:NJLint} is turned into a
Yukawa-interaction $h$ between quarks, mesons and diquarks. In such a
low-energy effective theory, the gluon fields are considered to be
integrated out. This approach is supported by the mass gap in QCD
which suppresses gluon fluctuations for low momentum scales. This is
most easily seen in the Landau gauge where the QCD mass gap translates
in a mass gap in the gluon propagator. Since the
mass gap of the gluon is of the order of $1\,\text{GeV}$, the
UV-cutoff of our model has to be chosen accordingly, i.e. $\Lambda
\approx 1$ GeV.

%%%%%%%%%%%%%%%%%%%%%%%%%%%%%%%%%%%
%%%%%%%%%%%%%%%%%%%%%%%%%%%%%%%%%%%

\subsection{Functional Renormalisation}
\label{sec:frg}
%
%%%%%%
%%%%%%%%%%%%
\begin{figure}[t]
\center
\includegraphics[width=0.82\columnwidth]{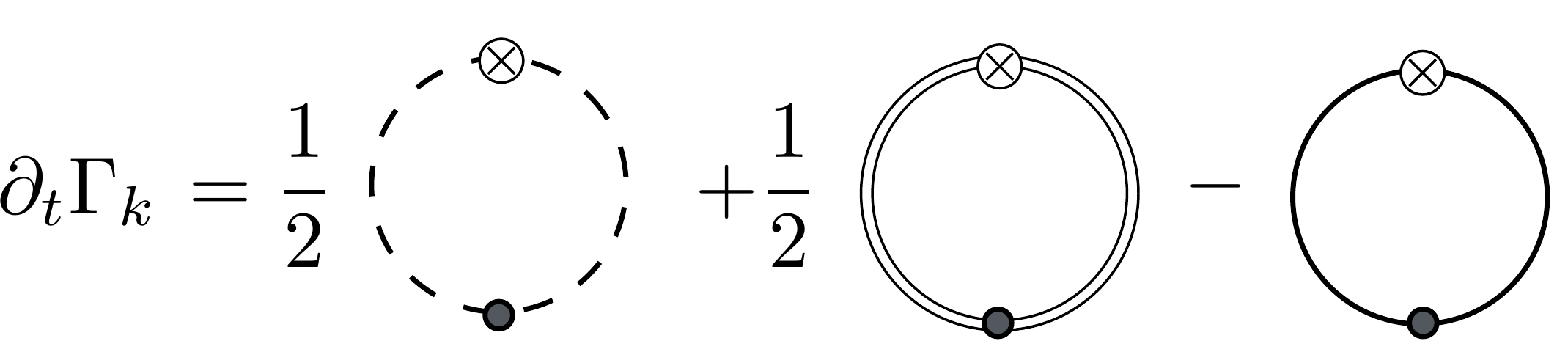}
\caption{RG flow of the effective action of the Quark-Meson-Diquark
  model (QMD). The meson, diquark an quark propagators are represented
  by dashed, double and solid lines respectively. The filled circles
  indicate that these are the full propagators. The crosses denote the
  regulator insertion $\partial_t R_k$.}
\label{fig:flowG}
\end{figure}
%%%%%%%%%%%%
%%%%%%
For our study of the phase diagram of the quark-meson-diquark (QMD)
model we use the functional renormalisation group (FRG) approach. The
FRG is formulated as a functional differential equation for the
scale-dependent effective action or free energy $\Gamma_k$. Its scale
dependence is governed by the Wetterich equation,
\cite{Wetterich:1992yh},
\begin{equation}\label{eq:flowG}
  \partial_t \Gamma_k[\Phi]=\frac{1}{2}\mathrm{Tr}\, 
  G_k[\phi] \,\partial_t R_k\,,\quad t=\log k/\Lambda\,,  
\end{equation}
with some reference scale $\Lambda$, typically the initial UV
scale, and $G_k[\Phi]$ is the full field-dependent propagator,  
\begin{align}\label{eq:Prop} 
 G_k[\phi]= \0{1}{\Gamma_k^{(2)}[\Phi]+R_k } \,, \qquad \Gamma^{(2)}_k=
\0{\delta^2\Gamma_k}{\delta\Phi^2}\,.
\end{align}
In \eq{eq:flowG}, \eq{eq:Prop}, the field $\Phi$ collects all
bosonic and fermionic degrees of freedom of the model.  The trace in
\eq{eq:flowG} contains a sum over internal indices with a
relative minus sign for fermions as well as a momentum
integration. The function $R_k=R_k(q)$ acts as a momentum-dependent
mass, suppressing infrared modes below the RG scale $k$ and leaving
the UV unchanged. The solution to \eq{eq:flowG} provides
a renormalisation group trajectory, interpolating between the
microscopic action $\Gamma_\Lambda$ at the ultraviolet scale $\Lambda$
and the full quantum effective action $\Gamma=\Gamma_{k\to 0}$.

The flow equation \eq{eq:flowG} is a functional
integro-differential equation which only in rare cases admits a full
solution. In most cases one has to resort to approximations of the
full effective action. For a discussion of suitable approximation
schemes and their convergence we refer to the extensive literature,
for QCD-related reviews see \cite{Litim:1998nf, Berges:2000ew,
  Polonyi:2001se, Pawlowski:2005xe, Gies:2006wv, Schaefer:2006sr,
  Rosten:2010vm, Braun:2011pp, vonSmekal:2012vx,Pawlowski:2014aha}.

In the present work we use the QMD model (\ref{eq:QMD}),
\cite{Strodthoff:2011tz}, as derived in the previous section. The flow
equation for the effective action is depicted diagrammatically in
\fig{fig:flowG}. It comprises quantum, thermal and density
fluctuations of quarks, mesons and diquarks. The approximation scheme
used in the present work is discussed in Sec.~\ref{sec:setup}.

%%%%%%%%%%%%%%%%%%%%%%%%%%%%%%%%%%%
%%%%%%%%%%%%%%%%%%%%%%%%%%%%%%%%%%%

\subsection{Silver Blaze Property}\label{sec:SilvBl}

At vanishing temperature the quark chemical potential~$\mu$ has to
exceed a critical value $\mu_c=m_{\Delta,\text{\tiny{pol}}}/2$, before
a finite density can be reached. Here, $m_{\Delta,\text{\tiny{pol}}}$
is the pole mass of the diquark, which is the lowest lying state with
non-vanishing baryon number.  For $\mu>\mu_c$, diquarks can be
generated leading to a finite density. For $\mu<\mu_c$ the density is
vanishing. The $\mu$ independence below the onset chemical potential
$\mu_c$ of the density translates to a trivial one for all correlation
functions in QC${}_2$D, see \eq{eq:SBp} and below, and also holds in
QCD with $\mu_c=m_{\rm baryon,\text{\tiny{pol}}}/3$. It is related to
the pole- and cut-structure of the correlation functions in momentum
space, and is called the {\it Silver Blaze property}
\cite{Cohen:2003kd}.

For 1PI correlation functions $\Gamma^{(n)}_k$ in the presence of
a regulator the Silver Blaze property reads
schematically
\begin{align}\label{eq:SBp}
  &\G^{(n)}_k(p_{1},\ldots,p_{n};\mu) =\G^{(n)}_k(\tilde p_{1},
  \ldots,\tilde p_n;0)\,,\quad \mu<\mu_c\,,
\end{align}
with 
\begin{align}\label{eq:tp}
  \G^{(n)}_k = \0{\delta^n\Gamma_k}{\delta\Phi^n} \,, \qquad \qquad
  \tilde p_i= ({p_i}_0+ i\,c_i\mu,\vec p_i)\,,
\end{align}
and $c_i/2$ is the baryon number of the field $\Phi_{i}$ in
QC${}_2$D. In QCD this translates into $c_i/3$ which reflects three
colors. \Eq{eq:SBp} requires regulators with
\begin{align}\label{eq:SBreg}
R_k(p;\mu)= R_k(\tilde p ;0)\,. 
\end{align}
Moreover, $R_k$ should not lower the pole mass of the lowest lying
state with non-vanishing baryon number. Note that \eq{eq:SBp} and
\eq{eq:SBreg} implies the same property for the propagator,
\begin{align}\label{eq:SBG}
G_k(p_1,p_2;\mu) = G_k(\tilde p_1,\tilde p_2;0)\,,\quad \mu<\mu_c\,. 
\end{align}
At $k\!=\!0$, \eq{eq:SBp} is the standard Silver-blaze property,
see~\cite{Marko:2014hea} for a recent lucid diagrammatic proof within
the 2PI-approach.

In summary \eq{eq:SBp} entails that below the critical or onset
chemical potential, $\mu_c$, the $\mu$-dependence of the $n$-point
functions is that of the frequency arguments, $p_i \!\to\! \tilde
p_i$. This also makes clear why the onset chemical potential is given
by the pole mass of the lowest lying state with non-vanishing baryon
number: the pole mass of, e.g. the diquark, is given by the condition
\begin{align}\label{eq:polemass} 
  \Gamma_{k,\Delta}^{(2)}(p^2=-m_{\Delta,\text{\tiny{pol}},\mu}^2;\mu)=0\,,\quad
  \quad \vec p=0\,,
\end{align} 
where $m_{\Delta,\text{\tiny{pol}},\mu}$ is the pole mass at finite
$\mu$. As $\Gamma_{k,\Delta}^{(2)}$ has the Silver Blaze property
\eq{eq:SBp}, this also holds for \mbox{$\Gamma_{k,\Delta}^{(2)}(\tilde
  p^2 = -m_{\Delta,\text{\tiny{pol}}}^2;0)$} with the pole mass
$m_{\Delta,\text{\tiny{pol}}}$ at vanishing $\mu$. \Eq{eq:polemass} is
fulfilled for $\tilde p^2 \!=\!  -m_{\Delta,\text{\tiny{pol}}}^2$
which implies $p^2 \!=\! 0$ and $\mu=\mu_c=
m_{\Delta,\text{\tiny{pol}}}/2$. Hence, onset chemical potential and
pole mass at vanishing $\mu$ are the same due to the Silver blaze
property. Note that this also implies that the pole mas at the
critical chemical potential vanishes,
$m_{\Delta,\text{\tiny{pol}},\mu_c}=0$. Moreover, at $\mu_c$ the
curvature mass, $\Gamma_{k,\Delta}^{(2)}(p^2=0)$, vanishes as well.

Note in this context that the curvature masses, e.g.\
$\Gamma_{k,\Delta}^{(2)}(p^2=0)$ are not necessarily identical with
the pole masses for all $\mu$. However, it has been observed in
\cite{Helmboldt:2014iya} within the quark-meson model that this
non-trivial relation holds well if fully momentum-dependent
propagators are taken into account. In turn it fails in the local
potential approximation. Additionally, it has been shown in
\cite{Helmboldt:2014iya} that the inclusion of wave function
renormalisations $Z_{k,\Phi}$ takes account of most of the momentum
dependence and hence the non-trivial relation between pole and
curvature mass is fulfilled with high accuracy. Thus, in the present
work we account for the non-trivial momentum dependence of the
two-point functions by the inclusion of running $Z_{q/\phi/\Delta}$.

The Silver blaze property also has important consequences for
approximation schemes of the effective action: in order to retain the
Silver Blaze property within a truncation, the frequency-dependence of
the $n$-point functions relating to fields with non-vanishing baryon
number has to be taken into account properly, see also
\cite{Fu:2015naa}. Alternatively one may expand the correlation
functions at $\tilde p_i=0$ with $\tilde p$ in \eq{eq:tp}. In the present
  truncation this concerns the wave function renormalisations of
  quarks and diquarks, the Yukawa couplings as well as the diquark
  couplings in the effective potential, and will be discussed in
  Sec.~\ref{sec:setup}.

\subsubsection{Silver Blaze Property from the functional RG}

We proceed by showing the Silver Blaze property~(\ref{eq:SBp}) with
the help of the FRG. The FRG allows for a particularly simple proof
due to its one-loop complete structure. The successive or iterative
nature of the flow gives access to an inductive proof, which
structurally works in the following way: we show that the flow of the
correlation functions at some scale $k$ has the Silver blaze
property~\eq{eq:SBp} if the correlation functions at this scale $k$
have it. Hence, if the Silver Blaze property holds at one scale, it
holds at every scale. In fact, due to asymptotic freedom, QC${}_2$D
approaches the classical action at asymptotically large scales. In
this case, the only momentum-dependent $n$-point function in QC${}_2$D
is the two-point function of the quarks for which (\ref{eq:SBp})
holds. This leads to a classical or initial effective action for the
QMD model which also has the Silver blaze property, see \eq{eq:QMD}.

It is left to show that Silver Blaze holds for the flows $\partial_t
\Gamma_k^{(n)}$ provided that it holds for all $\Gamma^{(n)}_k$. We
illustrate the structure of the proof at the example of the bosonic
two-point function in a vanishing background, $\Phi=0$. Furthermore,
for the sake of simplicity we drop the quark contribution,
effectively restricting ourselves to the purely bosonic sector of the
theory. In a vanishing background the correlation function is diagonal
in momentum space, $\Gamma^{(2)}_k(p_1,p_2;\mu)\simeq
\Gamma^{(2)}_k(p;\mu)$ times momentum conservation. This 
simplifies the flow, to wit  
\begin{align}\label{eq:SBflow1}
  \partial_t\G_k^{(2)}(p;\mu) =& -\frac{1}{2}\int_{q} 
 \G_k^{(4)}(p,q;\mu)   
\nonumber\\ 
  &\hspace{.6cm}\times
G_k(q;\mu) \,\partial_t R_k(q;\mu) \, G_k(q;\mu)\,, 
\end{align}
where we have used that the propagator as well as the four-point
function only depend on one and two momenta respectively. Now we
utilise the assumption that the correlation functions at the scale $k$
satisfy \eq{eq:SBp}, which requires regulators with \eq{eq:SBreg} and
implies \eq{eq:SBG}. In summary this leads to
\begin{align}\label{eq:SBflow2}
  \partial_t\G_k^{(2)}(p;\mu) =& -\frac{1}{2}\int_{\tilde q} 
 \G_k^{(4)}(\tilde p,\tilde q;0)   
\nonumber\\ 
  &\hspace{.6cm}\times
G_k(\tilde q;0) \,\partial_t R_k(\tilde q;0) \, G_k(\tilde q;0)\,, 
\end{align}
where we have shifted the frequency integration over real $q_0$ to one
over $q_0+i\,\mu$.  This is possible as long as the integrand has no
poles in the frequency integration contour given by $q_0, q_0+ i\,\mu$
and the connecting lines at $ \pm\infty + i y$ with $y\in [0,\mu
]$. For $\mu<\mu_c$ all poles are at larger imaginary frequency values
for all spatial momenta. Hence, the right hand side of \eq{eq:SBflow2}
only depends on $\mu$ via $\tilde p$ and we conclude
\begin{align}\label{eq:SBflow3}
  \partial_t\G_k^{(2)}(p;\mu) &= \partial_t\G_k^{(2)}(\tilde p;0) \,. 
\end{align}
The one line proof from \eq{eq:SBflow1} to \eq{eq:SBflow2} extends
trivially to the flow equation for $\Gamma_k^{(2)}$ of our complete
model including quark contributions, non-trivial background fields, and
all higher correlation functions, $\partial_t \Gamma^{(n)}_k$: all
diagrams have the same one-loop structure as \eq{eq:SBflow1}, the
difference only being a different number of vertices and
propagators. Hence, the substitution of the vertices and propagators
according to \eq{eq:SBp}, \eq{eq:SBG} works similarly. The shift of
all integrations $q_i \to \tilde q_i$ is possible by the lack of
poles between ${q_0}_i \to {q_0}_i+i\mu$ for $\mu<\mu_c$. This leads
us finally to
\begin{align}\label{eq:SBflow4}
  \partial_t\G_k^{(n)}(p_{1},...,p_n;\mu) &= \partial_t\G_k^{(n)}(
\tilde p_{1},...,\tilde p_{n};0) \,. 
\end{align}
We emphasise again that it is the one-loop structure of the flow that
simplifies the proof.

\section{Set-up and flow equations}\label{sec:setup}

Here we specify and discuss the truncation of the effective action we
use in this work. The flow of the cutoff-dependent parameters are then
extracted from the flow equation \eq{eq:flowG} for the
effective action by appropriate projections which is also discussed in detail. 

\subsection{Scale-dependent effective action}\label{sec:effac}
We promote all parameters of the QMD Lagrangian (\ref{eq:QMD})
to cutoff-scale dependent ones. In order to account for
fluctuation-induced modifications of the classical dispersion
relations, we also introduce running wave function renormalisations $Z_q$,
$Z_\phi$ and $Z_\Delta$ for the quarks, the mesons and the diquarks
respectively. The Pauli-G\"ursey symmetry at vanishing chemical
potential relates the meson and diquark Yukawa couplings to be
$h_\phi=h_\Delta = h$ as in \eq{eq:QMD}. However, at finite
$\mu$ this symmetry is broken which leads to manifestly different
interactions of diquarks and mesons. To accurately account for this
breaking we allow for different running Yukawa couplings $h_\phi \neq
h_\Delta$. Higher order scale-dependent interactions between mesons
and diquarks are taken into account by a scale-dependent effective
potential $V$.

Taking all this into account, the quark sector of our truncation,
i.e. the contribution to the effective action bilinear in the fermion
fields, reads
\begin{align}
  \Gamma_{q}&=\int_x\;Z_{q}\bar{q}\left( i\slashed{\pa}
    +i\sqrt{Z_{\sp\f}}h_{\sp\f}(\s+i\g_5\vec{\t}\cdot\vec{\p} )+
    i\gamma_0\m\right) q\nonumber\\
  &\quad +\int_x\;Z_{q}\sqrt{\frac{ Z_{\sp\D}}{2}}h_{\sp\D}\left(i\D^*
    q^T C \gamma_5 \tau_2 t_2 q-{\rm
      h.c.}\right)\,.\label{eq:quarkaction}
\end{align}
It contains the kinetic term for the quarks as well as the Yukawa
interaction terms. The prefactor $\sqrt{Z_\phi}$ in the Yukawa
interaction is introduced in order to guarantee RG-invariance of the
Yukawa couplings: the RG-scaling is carried by the factor $Z_q
\sqrt{Z_\phi}$. Note however, that RG-invariance does not imply cutoff
independence.  The kinetic terms of the mesons and the diquarks are
\begin{align}\label{eq:MDaction}
  \Gamma_{\sp\phi}&=\int_x\left\lbrace \frac{Z_{\sp\f}}{2} \left[
      \left(\pa_\m \vec{\p}\right) ^2+ \left(\pa_\m \s
      \right) ^2\right]-\sqrt{Z_{\sp\f}}c\s\right\rbrace\,,\nonumber\\
\Gamma_{\sp\Delta}&=\hspace{-0.1cm}\int_x{Z_{\sp \D}}\left\lbrace
    \left(\partial_\tau-2\mu\right)\D^*\left(\partial_\tau+2\mu\right)\D
    +\left| \partial_i\D\right|^2\right\rbrace.
\end{align}
Note that we have included the explicit chiral symmetry breaking
parameter $c$ to the mesonic contribution. The derivative term of the
diquarks includes the chemical potential and the factor $\pm2$
indicates the corresponding charge of the (anti-)diquark. We also have
rescaled the diquark fields with a factor $\sqrt{2}$ in comparison to
\eq{eq:QMD}, in order to obtain the conventional action of a complex
scalar field. In our Euclidean setting the time axis is compactified
on a torus with circumference $\beta = 1/T $ at finite temperature. In
case the $SU(4)$ flavor symmetry is not violated, the relations
$Z_\f=Z_{\sp\D}$ and $h_{\sp \f}=h_{\sp \D}$ hold. In summary, the
complete truncation of the full effective action reads,
\be \Gamma[\Phi]=\Gamma_{q}+\Gamma_{\sp\phi}
+\Gamma_{\sp\Delta}+\int_x \;V(\bar\varphi)\,,\label{eq:QMDaction} \ee
where $\F=\left( \vec{\p},\s,\D,\D^*,q,\bar{q}\right)$, and the
effective potential $V(\bar\varphi)$ contains the multi-meson diquark
interactions. It is discussed in more detail in the next section.

%%%%%%%%%%%%%%%%%%%%%%%%%%%%%%%%%%%
%%%%%%%%%%%%%%%%%%%%%%%%%%%%%%%%%%%

\subsection{Effective Potential}\label{sec:effpot}

At vanishing chemical potential two-flavor QC${}_2$D exhibits the
$SU(4) \simeq SO(6)$ Pauli-G\"ursey flavor symmetry. At finite $\mu$
this is broken to $SO(4)\!\times\!SO(2)$. $SO(4)$ is related to the
$SU(2)_L\!\times\!SU(2)_R$ chiral symmetry and $SO(2)$ is related to
the $U(1)_B$ baryon number symmetry. Hence, the effective potential
can be considered as a function of two invariants. Taking the explicit
symmetry breaking into account, we have
\bea U(\r_{\sp \f},\r_{\sp\D},\s)=V(\r_{\sp
  \f},\r_{\sp\D}) -4\m^2 \r_{\sp\D} - c
\sqrt{Z_{\sp\f}}\s \,, \label{eq:potrnc} \eea 
where 
\bea \r_{\sp
  \f}=\frac{Z_{\sp \f}}{2}(\vec{\p}^2+\s^2) \, ,\quad
\r_{\sp\D}=
\frac{Z_{\sp\D}}{2}(\D_1^2+\D_2^2)\, ,\label{eq:rho} 
\eea 
are the invariants of the chiral and baryon number symmetry
respectively. The field invariant $\r_{\sp\D}$ is given in the real
representation of the diquarks, which relate to the complex
representation by $\D=\frac{1}{\sqrt{2}}(\D_{1}+i\D_{2})$ and
$\D^{*}=\frac{1}{\sqrt{2}}(\D_{1}-i\D_{2})$. The explicit current
quark mass, which directly relates to the explicit symmetry breaking
$c$, gives rise to a smooth chiral crossover. On the other hand, a
rising chemical potential will lead to a second order phase transition
to the BEC phase.

A non-vanishing vacuum expectation value $\kappa_{\sp\f}$ of
$\r_{\sp\f}$ signals chiral symmetry breaking and a non-vanishing
vacuum expectation value $\kappa_{\sp\D}$ of $\r_{\sp\D}$ signals the
breaking of baryon number conservation. They are given by the
stationary solution of the effective potential $U$: $\partial_\phi
U=\partial_\Delta U=0$. With \eq{eq:potrnc} this leads to stationary
conditions for the effective potential $V$. For the radial directions
$\sigma$ and $\Delta_1$ these equations read 
\begin{eqnarray}
\frac{\pa V}{\pa\r_{\sp \f} }
    \bigg\vert_{\vec \r=\vec \k}
    = \frac{ c}{\sqrt{2\k_{\sp \f}}}\,,\qquad \qquad 
\frac{\pa V}{\pa\r_{\sp \D} }\bigg
  \vert_{\vec \r=\vec \k}= 4\m^2 \,. \label{eq:min}
\end{eqnarray}
where $\vec \r =(\r_{\sp \f},\r_{\sp \D})$ and $\vec \k =(\k_{\sp
  \f},\k_{\sp \D})$. The solution $\kappa$ of \eq{eq:min} are the
order parameters for the chiral and the BEC phase respectively.

%%%%%%%%%%%%%%%%%%%%%%%%%%%%%%%%%%%

\subsection{Flow of the Effective Potential}\label{ssec:FleqQMD}
The flow of the effective potential $U(\rho_\phi,\rho_\Delta)$ is that
of the effective action, \eq{eq:flowG}, depicted in \fig{fig:flowG},
evaluated at constant bosonic fields and vanishing quarks. Then all
quark terms and all derivative terms in the effective action vanish
and the left hand side of the flow equation, \eq{eq:flowG}, reduces to
$\partial_t U$. With the flat 3d regulators specified in
App.~\ref{app:Fleq}, \eq{eq:regs}, the spatial momentum integration is
done readily. This leads to
\begin{align}
  \pa_t {U}(\r_{\sp \f},\r_{\sp\D})&= \frac{k^5T}{6\p^2}\sum_{n \in
    \mathds{Z}} \Biggl[ \left(1-\frac{\h_{\sp \f}}{5} \right)\left[ 3
    G_\p(k^2)+G_\s(k^2)\right]
  \nonumber \\
  +2 \Big(1-&\frac{\h_{\sp \D}}{5} \Big) G_{\sp \D}^+(k^2)+4 N_c N_f
  \left(1-\frac{\h_{q}}{4} \right)A_+
  (k^2)\Biggr]\,, \label{eq:dtUgen}
\end{align}
where the boson and fermion propagators $G$ and $A$ are given in
App.~\ref{app:prop}, and anomalous dimensions are defined as $
\h_{\F_i}=-\pa_tZ_{\F_i}/Z_{\F_i}$. The Matsubara summation can be
carried out analytically, but due to the excessive length of the
resulting equation we do not show it here.

\begin{figure}[t]
\begin{eqnarray*}
\raisebox{-10ex}{\includegraphics[height=18ex]{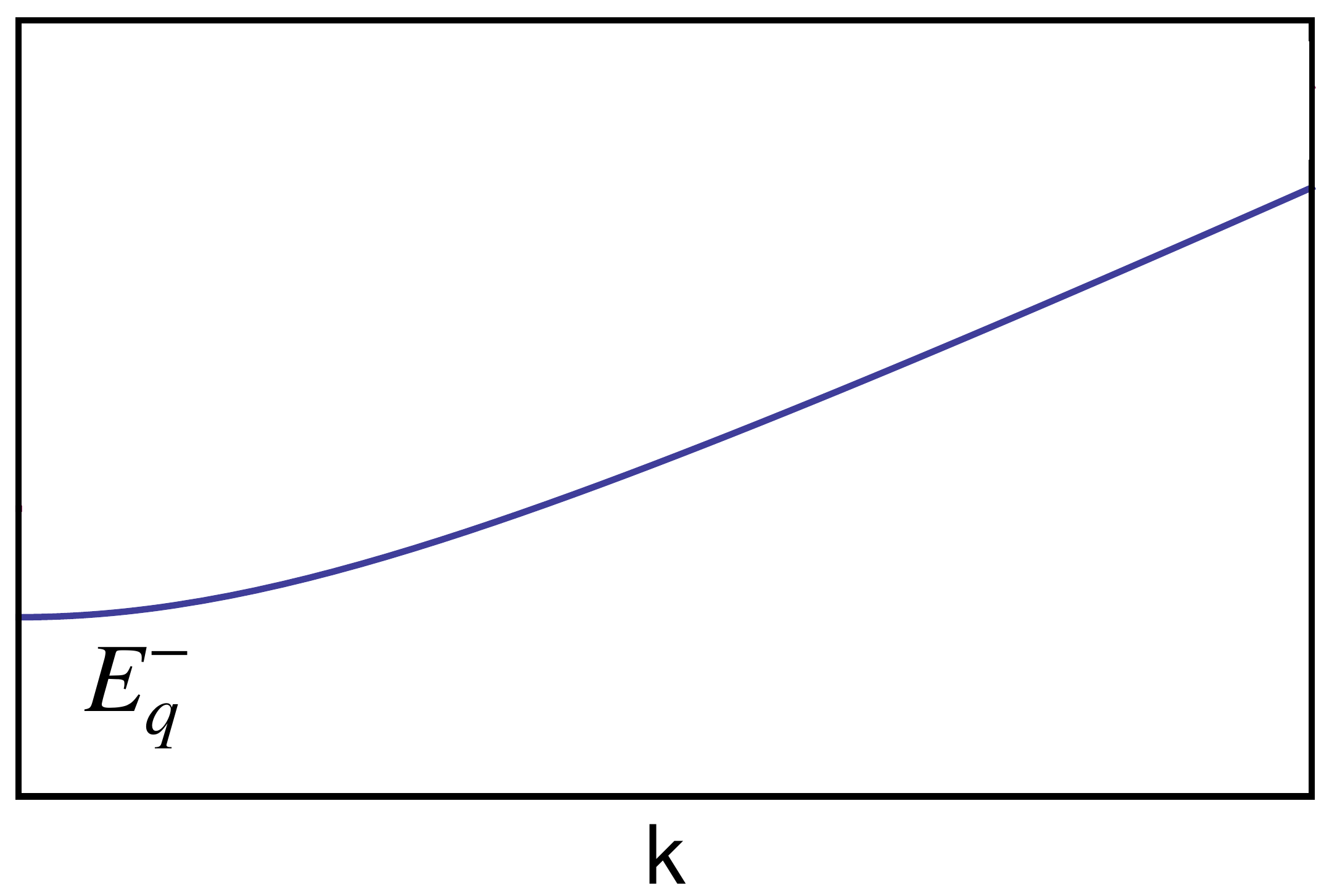}}\xrightarrow{ k_F>0}
\raisebox{-10ex}{\includegraphics[height=18ex]{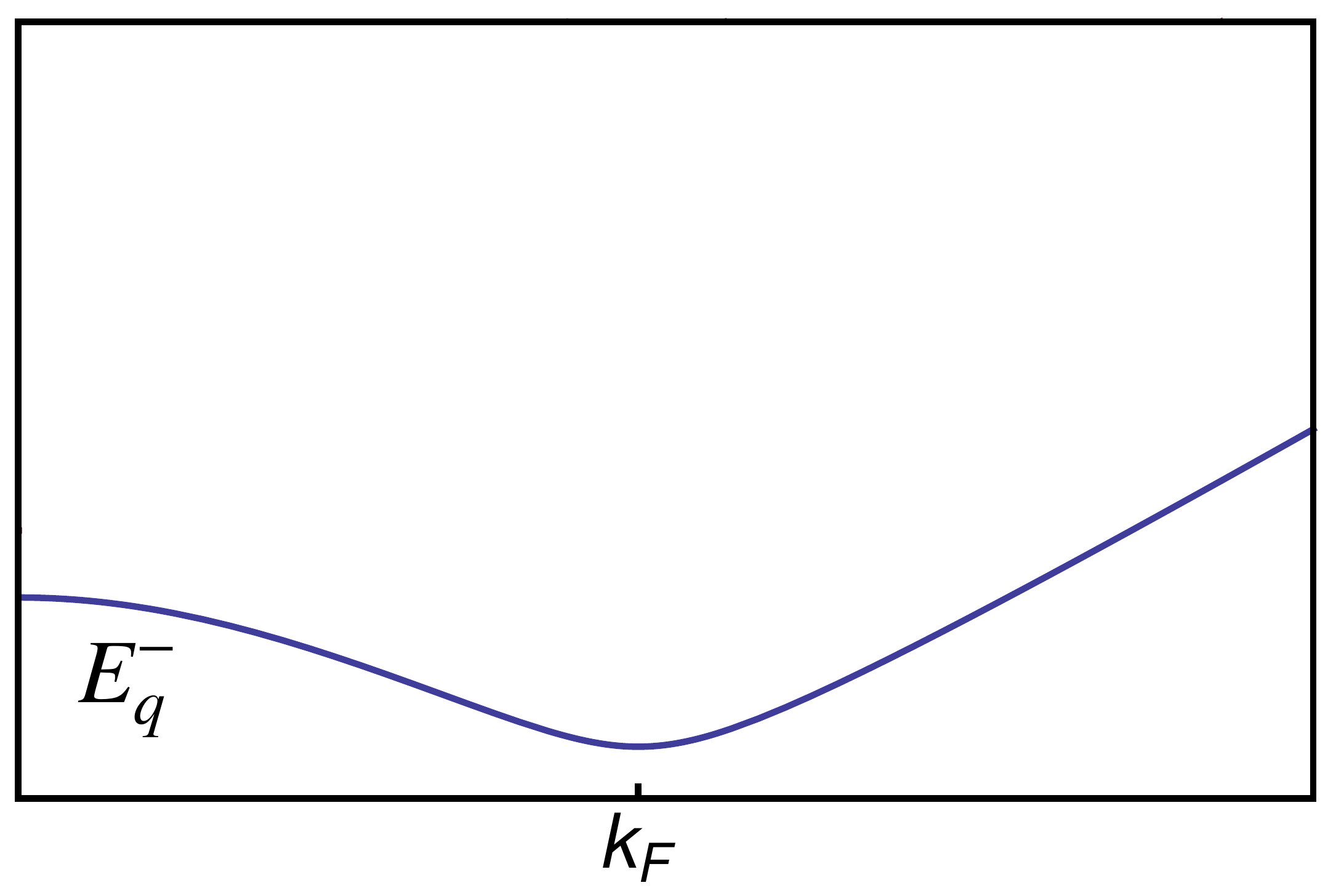}}
\end{eqnarray*}
\caption{On the left-hand side we see the qualitative behaviour of the quark
  energy dispersion for the case $k_{F}<0$. It is a monotonically
  rising functions with $k$. On the right-hand side we have the case $
  k_F>0$. The dispersion develops a minimum at the Fermi surface,
  which indicates the formation of quark Cooper
  pairs.}
\label{fig:BecBsc}
\end{figure}

The right hand side of the flow equation \eq{eq:dtUgen} depends on the
energy dispersion relations of the fields, that can be read-off from
the corresponding propagators in App.~\ref{app:prop}. The onset of the
diquark phase is reflected in these energy dispersion relations. In
particular, the energy dispersion of the quarks is
\begin{align}  \label{eq:modgap}
E_q^{\pm}=\sqrt{\left(\e_q\pm \m \right) ^2+2h_{\sp \D}\r_{\sp \D}}\,,\;\; 
\e_q=\sqrt{k^2+2h_{\sp \f}\r_{\sp \f}}\,.
\end{align}
At finite chemical potential a Fermi surface occurs and we define the
corresponding Fermi energy as $k^{2}_{F} \!=\! \m^{2} \!-\!
m_{q}^{2}$, where $m_{q}\!=\!\sqrt{2h_{\sp\f}\r_{\sp\f}}$. The quark
energy dispersion $E_q^-$ for the case $k_F\!<\!0$ in the phase with
$\rho_\Delta \!\neq\! 0$ is shown in the left plot of
Fig.~\ref{fig:BecBsc}. The minimum of the energy dispersion at
vanishing momentum $k\!=\!0$ indicates that the diquark condensate is
of the BEC type. When the chemical potential exceeds the quark mass,
i.e. $k_F\!>\!0$, the quark energy dispersion develops a minimum at
non-vanishing momentum scale $k_\text{min} \!=\! k_F$. This is shown
in the right plot of Fig.~\ref{fig:BecBsc}. This is typical for a
BCS-like ground state where the quarks form Cooper pairs. Hence,
$m_{q} \!=\! \mu$ can be used as an indication for the BEC-BCS
crossover in the present case \cite{He:2010nb}.

%%%%%
%%%%%%%%%%%%%
%%%%%%%%%%%%
%%%%%

%%%%%%%%%%%%%%%%%%%%%%%%%%%%%%%%%%%
%%%%%%%%%%%%%%%%%%%%%%%%%%%%%%%%%%%

\subsection{Flow of the Meson-Diquark Couplings}\label{ssec:flowofparam}

We expand the effective potential in the invariants $\rho_\phi,
\rho_\Delta$ about its flowing minimum $\vec \kappa$, see
\eq{eq:min}. For related recent FRG-applications on such a
two-dimensional Taylor expansion for the effective potential see
e.g.~\cite{eichhorn2013multicritical,classen2015competition}. Alternative
evaluation methods of the flow of the effective potential are a grid
of field values \cite{Strodthoff:2011tz} or an expansion about a fixed
background \cite{Pawlowski:2014zaa}. In such a two-dimensional Taylor
expansion the potential reads
\begin{align}
  V(\r_{\sp\f},\r_{\sp\D})&= \sum_{\substack{n,m=0\\ n+m<N}}
  \frac{\l_{n,m}}{n!m!}(\r_{\sp\f}- \k_{\sp\f})^n(\r_{\sp\D}-
  \k_{\sp\D})^m\,,\label{eq:2dtaylor}
\end{align}
where the linear couplings are related to mass parameters,
\begin{align}
  \lambda_{1,0}= m_\phi^2\,,\qquad \qquad \lambda_{0,1}=
  m_\Delta^2\,,\label{eq:masses}
\end{align} 
and $\lambda_{0,0}$ determines the value of the potential $V$ in the
minimum, $V(\vec \kappa)= \lambda_{0,0}$. This is relevant
for the thermodynamics of the system, as the pressure is related to
$V_T-V_{T=0}$.  In the present work we do not discuss the
thermodynamical properties of QC${}_2$D, and drop the constant part.

Physically, the expansion in \eq{eq:2dtaylor} constitutes an expansion
of the effective potential in terms of multi-scattering processes of
mesons and diquarks.  By solving \eq{eq:min} with this
parameterisation, we obtain the order parameters of chiral and baryon
number symmetry breaking.  For the chiral order parameter we find in all phases,
\bea \k_{\sp \f}=\frac{1}{2}\left(\frac{c}{m_{\sp\f}^2}\right)^2\,,
\label{eq:chirMin} \eea
For diquark condensation we distinguish two cases. The normal phase is
defined by $\k_{\sp\D} \!=\! 0$. The BEC regime is characterised by a
non-vanishing diquark condensate $\k_{\sp\D} \!\neq\! 0$. In this case
\eq{eq:min} implies $m_{\sp\D} \!=\! 2\m$.

We have introduced an explicit symmetry breaking $-c\sigma$ which is
directly linked to non-vanishing current quark masses. Since the flow
equation (\ref{eq:flowG}) is of second order in field
derivatives, we infer from \eq{eq:potrnc},  
\be \frac{\pa}{\pa t}\left(c\sqrt{Z_{\sp
      \f}}\right)=0\label{eq:noCflow}\;\;\;\Leftrightarrow \;\;\;\pa_t
c=\frac{\h_\f}{2} c\,.\label{eq:flowofc} \ee
We want to emphasise that this term merely shifts the minimum of the
effective potential without influencing the RG-flow of the effective
potential itself. However, since we expand about this minimum, $c$
enters the flow equations for the couplings via the expansion point.

The cutoff derivative of the effective potential at fixed
$\phi,\Delta$ in the parameterisation \eq{eq:2dtaylor} hits
$\lambda_{n,m}$, the flowing minimum $\kappa$ as well as the
renormalisation factors in $\vec \rho$. Therefore the
$\rho$-derivatives of $\partial_t U$ read
\begin{align}\nonumber 
  & \dot \l_{n,m} -\lambda_{n,m}\left( n\,\eta_\phi+m\,\eta_\Delta\right) -
    \l_{n+1,m}\left( \dot \k_{\sp\f} +\eta_\phi\kappa_\phi \right)
    \\[2ex]\nonumber 
    &-\lambda_{n,m+1}\left(\dot \kappa_\Delta +\eta_\Delta\kappa_\Delta  
 -4\mu^{2}\d_{n0}\d_{m1}\right)\\[2ex]
    = &
    \frac{\pa^{n}}{\pa\r_{\sp\f}^{n}}\frac{\pa^{m}}{\pa\r_{\sp\D}^{m}}
    \dot U \bigg\lvert_{\vec \r=\vec \k}\,,\label{eq:flowCoupl}
\end{align}
where $\dot U$ is given by the flow equation \eq{eq:dtUgen}. The
mesonic mass parameter $m_\phi^2=\lambda_{1,0}$ is connected to the
expansion point $\kappa_\phi$ via \eq{eq:chirMin}, and its flow is given by
\begin{align}
  \dot m_\phi^2=- \frac{c}{(2\k_{\sp\f})^{3/2}}\left( \dot
    \kappa_\phi- \eta_\phi\kappa_\phi \right) \,.\label{eq:dtm}
\end{align}
Furthermore, the cutoff-dependence of the diquark condensate and mass
parameter in the different phases can be inferred from \eq{eq:min} and
we find
\begin{align}
  &\text{Normal regime:}& \dot\k_{\sp\D} &= \k_{\sp\D} =0\;,
  \nonumber \\
  &\text{BEC regime:}& \dot m_\Delta^2&=0\;, \qquad
  m_\Delta^2=(2\m)^{2}\;. \label{eq:phases}
\end{align}
Finally, we discuss the implications of the Silver Blaze property
\eq{eq:SBp}: the couplings $\lambda_{n,m}$ carry baryon number $2m$.
Hence, at vanishing temperature and below the onset chemical potential
$\mu_c$ they are functions of $\tilde p_i$,
see~\eq{eq:SBp}. Accordingly, Silver blaze is violated with an
expansion at $p=0$ as is done for the flow of the effective potential.
Indeed it relates to a $\mu$-dependent expansion point $\tilde p =
i\,\mu$. We emphasise that this has nothing to do with the Taylor
expansion used here, but Silver blaze breaking is introduced by
considering constant diquark fields, leading to
\eq{eq:dtUgen}. Indeed, in contradistinction to a grid solution of
\eq{eq:dtUgen} the present Taylor expansion gives us the possibility
to even cure the mild Silver blaze breaking. First, we can evaluate
the flow of the couplings $\lambda_{n,m}$ at $\tilde p_i=0$. This
possibility will be explored in a subsequent work. Second, we may
enforce the Pauli-G\"ursey symmetry also in the presence of a
non-vanishing chemical potential and only expand the effective
potential in the meson field. This possibility is explored in the
App.~\ref{app:1dt} and neglects the physics effects of the breaking of
the Pauli-G\"ursey symmetry. Given the good convergence of the
derivative expansion this is less trustworthy.

%%%%%
%%%%%%%%%%%%%%
\begin{table*}[t]
\begin{center}
\begin{tabular}{|c||c|c|c|c|c|c|c|}
  \hline
  &$\Lambda$  [MeV]   &$\langle \s \rangle_{\sp\Lambda} $ [MeV]  & $m_{\sp \f,\sp\Lambda}$  
  [MeV]  & $\l_{2,0,\sp\Lambda}$  & $h_{\sp \f,\sp\Lambda}$    & $m_{\p}$   [MeV]  & $2 \m_c$  
  [MeV]    \\\hline\hline
  2d Taylor${}^\prime$ & 900   & 2.28 & 1135 & 89.0& 6.43 & 143 &  138  \\\hline
  2d Taylor & 900   & 4.50 & 650 & 7.0 & 4.80 & 158 & 139    \\\hline
  2d Taylor \cite{Strodthoff:2011tz} & 900   & 39.94 & 247 & 76.3 & 4.80  & 180 & 143  
  \\\hline
  $f_\pi=93$ MeV & 900   & 7.55 & 566 & 45.8& 4.14  & 138&  132 \\\hline
\end{tabular}
\end{center}
\caption{Initial conditions for the UV action $\G_{k=\Lambda}$, resulting 
  pion masses  $m_\p$ in the vacuum and the corresponding critical 
  chemical potential for the onset of diquark condensation at $T=0$.  
  The prime denotes that running Yukawa couplings and wave function 
  renormalisations are included. In the third row are the initial conditions 
  used in \cite{Strodthoff:2011tz}, where an LPA was solved on a 2d grid. 
  The resulting $m_\p$ and $2\m_c$ are shown for our 2d Taylor method, 
  and are close to what was found in \cite{Strodthoff:2011tz}.}
\label{tab:initial}
\end{table*}
%%%%%%%%%%%%%
%%%%%

In summary, the derivative expansion at vanishing momenta $p_i=0$ employed in
this work is based on the assumption of a mild $\tilde
p$-dependence. Hence we expect the breaking of Silver blaze to be
minimal, and we shall prove below that this is indeed the case.

\subsection{Anomalous Dimensions}\label{sec:anomalous}

The cutoff-dependent wave function
renormalisations enter the flow equations through the
corresponding anomalous dimensions $\eta_\Phi=-\partial_t
Z_\Phi/Z_\Phi$. In the spirit of the derivative expansion used here we
evaluate them at vanishing momentum. For the bosonic anomalous
dimensions this entails that the evaluation point is at $p^2 \!=\!0$.
Following our discussion in Sec.~\ref{sec:SilvBl} this results in a
manifestly $\mu$-dependent bosonic wave function renormalisations
below $\mu_c$. However, we demonstrate in App.~\ref{sec:blpa} that the
resulting violation of the Silver Blaze property is minor. In summary,
we define the bosonic anomalous dimensions via 
\bea
\h_{\varphi_i}&=&-\frac{1}{Z_{\vf_i}}\frac{\pa}{\pa
  \vec{p}^{\,2}}\dot{{\G}}^{}_{\vf_i\vf_i}\Big\lvert_{p=0}\,,
\label{eq:BosonicAno}
\eea 
where we use $\varphi_1 \!=\! \p_1$ for the mesons and
\mbox{$\varphi_6 \!=\! \D_2$} for the diquarks.  For explicit flow
equations see App. \ref{app:Fleq}.

Now we turn to the fermionic anomalous dimension, which we define via
\bea \h_q=-\frac{1}{3N_fN_c Z_q}\, \text{Re}\!\left[ \frac{\pa}{\pa
    \vec p}\cdot\tr \left(\vec \g\, \dot{{\G}}^{}_{\bar{\J}\J}\right)
  \Big\lvert_{p=p_{\text{min}}}\right]\,.
\label{eq:ZQproj}
\eea $\J$ represents the fermions in Nambu-Gorkov formalism defined in
App.~\ref{app:NGP}. The trace acts in all fermionic subspaces,
i.e. color-space, spinor-space and flavor space. The explicit flow
equation is shown in App.~\ref{app:Fleq}. Since fermions do not have a
vanishing Matsubara mode, we define the quark anomalous dimension at
the minimal momentum $p_{\text{min}} \!=\! (\p T,\vec{0})$ at finite
temperature \cite{Pawlowski:2014zaa}. This procedure guarantees that
fermion and boson propagators always carry the correct Matsubara
frequency in the loops. However, as a result, $Z_{q}$ is a function of
$(\p T \!-\! i\m)$ which, in addition to the explicit
$\mu$-dependence, renders the anomalous dimension complex valued. In
order to keep the effective action real, we project out the real part
in \eq{eq:ZQproj}. As for the bosonic wave function
renormalisations, we we demonstrate in App.~\ref{sec:blpa} that the
violation of the Silver Blaze property which results from the present
momentum-independent approximation of $Z_{q}$ is very small.

\subsection{Flow of the Yukawa Couplings}\label{ssec:Yukawa}

We can extract the running of the meson and diquark Yukawa coupling
from the flow of the quark two-point function similarly to
\cite{Pawlowski:2014zaa} and find
\begin{align}\label{eq:hproj}
  \begin{split} {\pa}_t h_{\sp\f} &=
    \left(\h_{q}+\frac{\h_{\sp\f}}{2}\right)h_{\sp\f}- 
    \frac{1}{4N_fN_c}\, \text{Re} \left[
      \tr\,\0{i\,\dot{{\G}}^{}_{\bar{\J}\J}}{ Z_{q} Z^{1/2}_\phi
        \sigma}\right] \, ,
    \\[2ex]
    {\pa}_t h_{\sp\D}&=
    \left(\h_{q}+\frac{\h_{\sp\D}}{2}\right)h_{\sp\D}+
    \frac{1}{4N_fN_c} \text{Re} \left[ \tr\,
      \hat{P}\,\0{\dot{{\G}}^{}_{\bar{\J}\J}}{ Z_{q} Z^{1/2}_\Delta
        \Delta} \right]\,. 
\end{split}
\end{align}
Here, $\hat{P} $ is the projection matrix on the diquark term in the
inverse quark propagator, see App.~\ref{app:prop},
\eq{eq:hatp}. \Eq{eq:hproj} provide flows for field and
momentum-dependent couplings. In the present work we evaluate the
flows on the flowing minimum $\vec \kappa$ as well as vanishing
momentum. The explicit expressions for the resulting resulting flow
equations are deferred to App.~\ref{app:prop}. The flow of both
couplings is complex at finite temperatures similarly as that of the
anomalous dimensions discussed in the previous section. Our projection
\eq{eq:hproj} is chosen such, that this does not lead to a complex
action. Moreover, the Silver Blaze violation related to the expansion
at vanishing momentum is small, see App.~\ref{sec:blpa}.

%%%%%%%%%%%%%%%%%%%%%%%%%%%%%%%%%%%
%%%%%%%%%%%%%%%%%%%%%%%%%%%%%%%%%%%

%%%%%
%%%%%%%%%%%%%%%
 \begin{figure*}[t]
    \includegraphics[scale=0.52, trim=0.7cm 0 0 0]{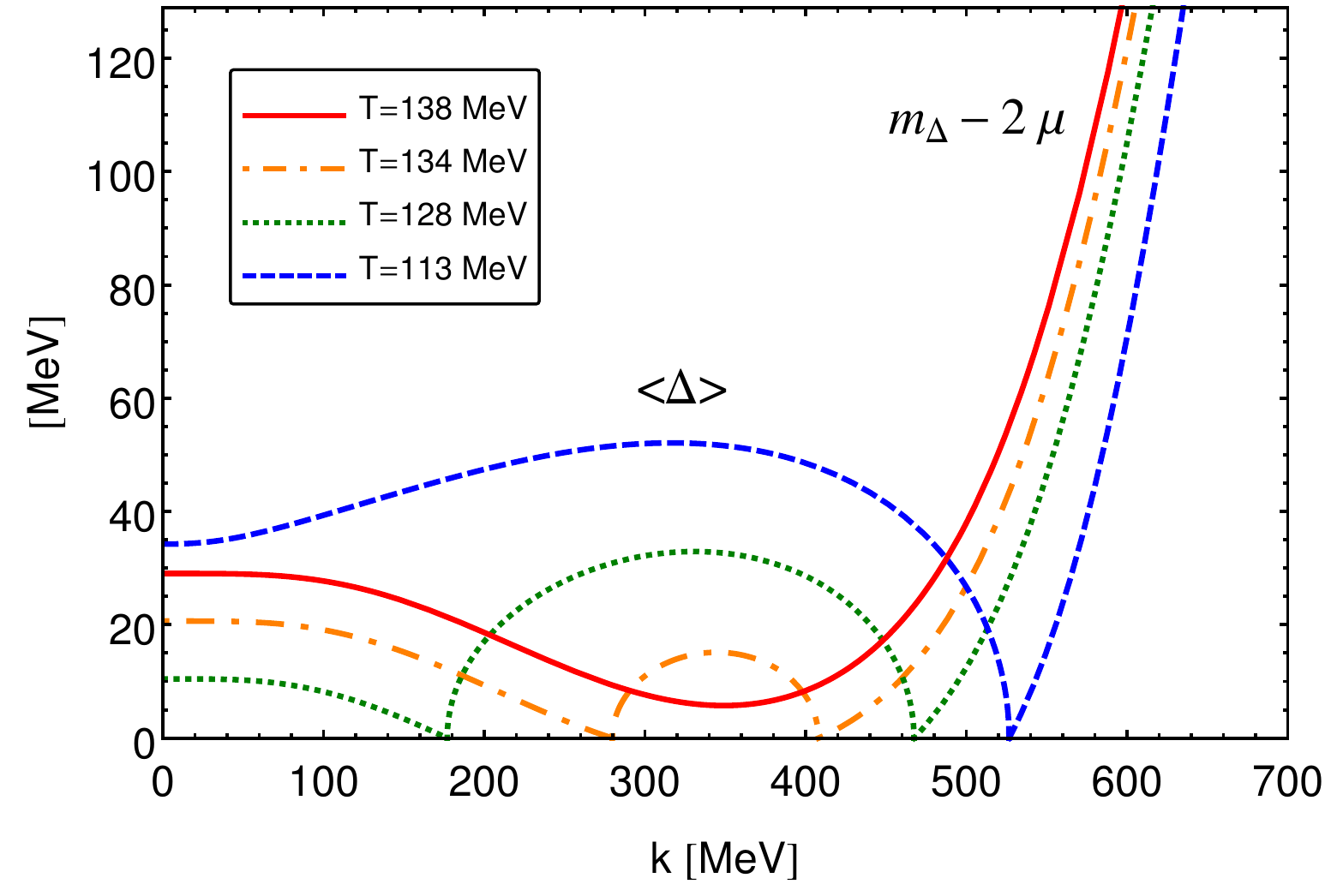}
    \includegraphics[scale=0.52, trim=-1.4cm 0 0 0]{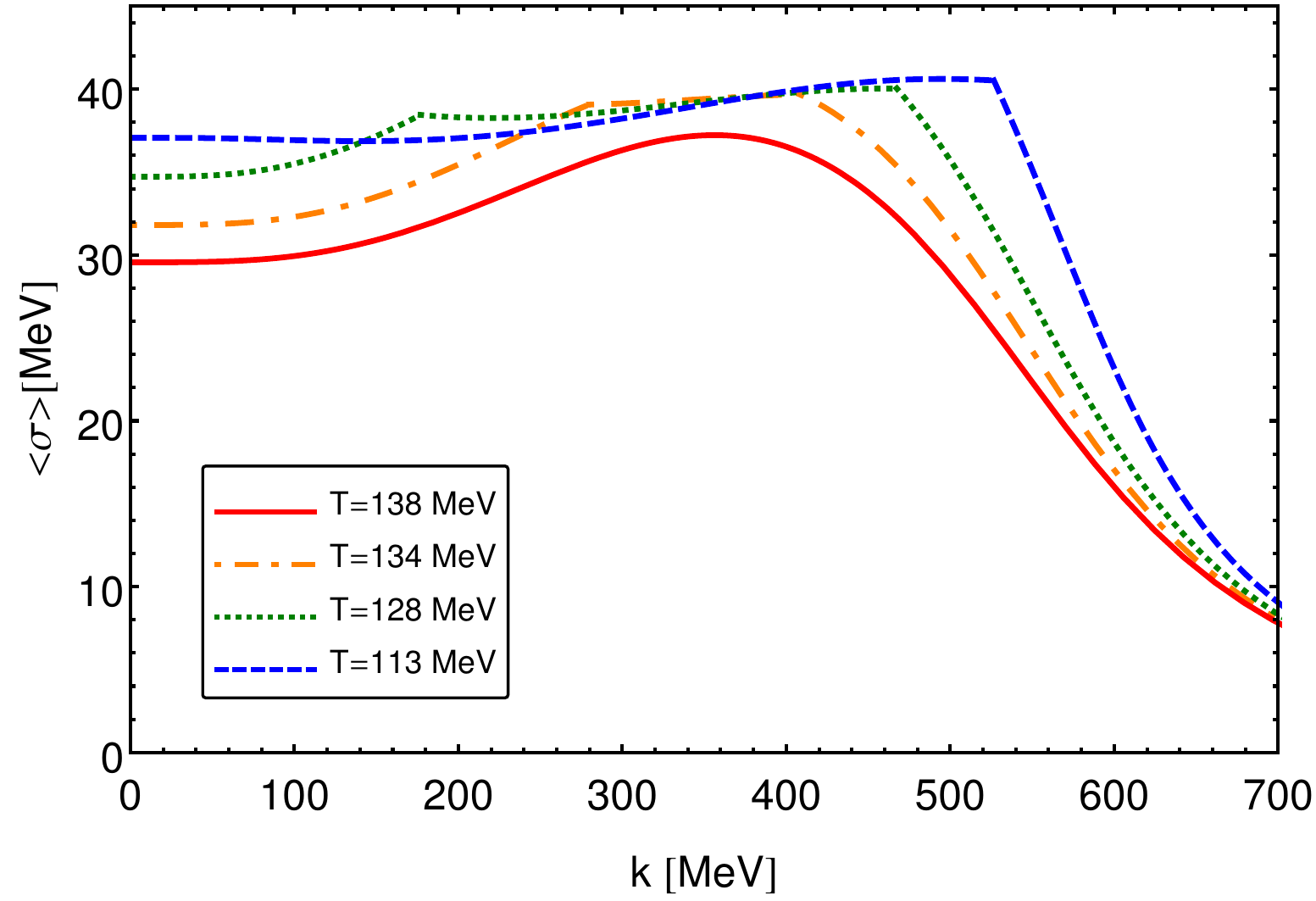}
    \caption{Flow of the diquark mass and condensate $\langle\D\rangle
      = \sqrt{2\k_{\sp \D}}$ (left panel), as well as the chiral
      condensate $\langle\s\rangle = \sqrt{2\k_{\sp \f}}$ (right
      panel) for \mbox{$\mu=100$ MeV} and different temperatures
      within the 2d Taylor${}^\prime$ truncation. The critical
      temperature for this choice of $\mu$ is $T_c \approx
      125\,\text{MeV}$. We see that the system exhibits finite domains
      of non-vanishing diquark condensate close to the transition to
      the BEC phase.}
\label{fig:Precondflow}
    \end{figure*}
%%%%%%%%%%%%%
%%%%%

%%%
%%%%%%%%%
\begin{figure}[t]
\centering	
\includegraphics[scale=0.5]{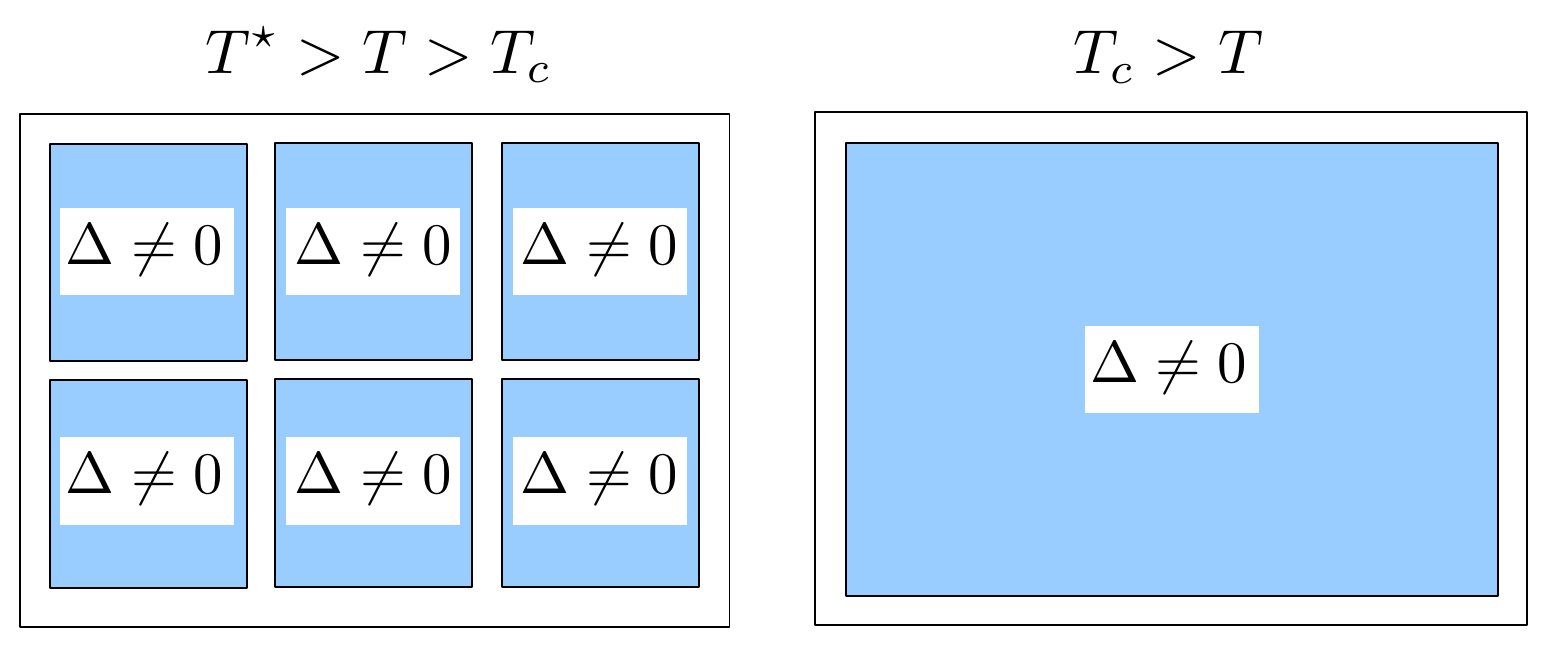}
\caption{Illustration of the pre-condensation effect. At temperatures
  close to $T_{c}$ local domains of condensation form (left
  figure). Below $T_{c}$ the whole volume is filled with a condensate
  (right figure). $T^\star$ is the temperature where pre-condensation
  sets in. It marks the outer edge of the shaded blue area in the
  right plot of Fig.~\ref{fig:QC2DPhasenDiagramm}.}
\label{fig:domains}       
\end{figure}
%%%%%%%%%
%%%

\subsection{Initial Conditions}\label{sec:initconds}

In Tab.~\ref{tab:initial} the initial conditions are displayed for
different truncations. The full truncation used in the present work is
denoted as 2d Taylor${}^\prime$. It has three running wave function
renormalisations $Z_{\sp \f}$, $Z_{\sp \D}$, $Z_{q}$, and two running
Yukawa couplings $h_{\sp \f}$, $h_{\sp \D}$. The Taylor expansion is
done up to the order $N\!=\!5$ in the $\r's$. In the UV we start with
an $SO(6)$ symmetric potential.

The initial conditions for the first two cases in the table are tuned
such that we obtain an $\sqrt{N_c}$-scaled pion decay constant of
$f_\p= \langle\s\rangle\bigr|_{k=0}=76$ MeV in the vacuum instead of
the usual $f_\p\simeq93$ MeV. Further conditions are that the vacuum
quark mass comes is $m_q\simeq360$ MeV in the IR, and that the onset
of the diquark condensation should be at $2\mu_c\simeq138$ MeV.

Note that physically sensible initial conditions in the chirally
symmetric phase feature small coupling strengths and large masses of
the bound states, which reflect their nature as auxiliary fields in
this phase. Hence, in addition to the initial conditions for the LPA
taken from \cite{Strodthoff:2011tz}, we have chosen the set shown in
the second row of Tab.~\ref{tab:initial}. We see that this choice
already reduces the difference between curvature mass of the pion and
onset chemical potential by about 50\% as compared to the initial
values taken from the reference.

For completeness, we also show initial conditions which yield
$f_\pi\simeq 93$ MeV and $m_q\simeq340$ MeV in the vacuum in the
fourth row of Tab.~\ref{tab:initial}. The main results of this work
are obtained with the 2d Taylor${}^\prime$ initial conditions.

\section{Results}\label{sec:RES}

%%%
%%%%%%%%%%%
 \begin{figure*}[t]
    \includegraphics[scale=0.4, trim=0.7cm 0 0 0]{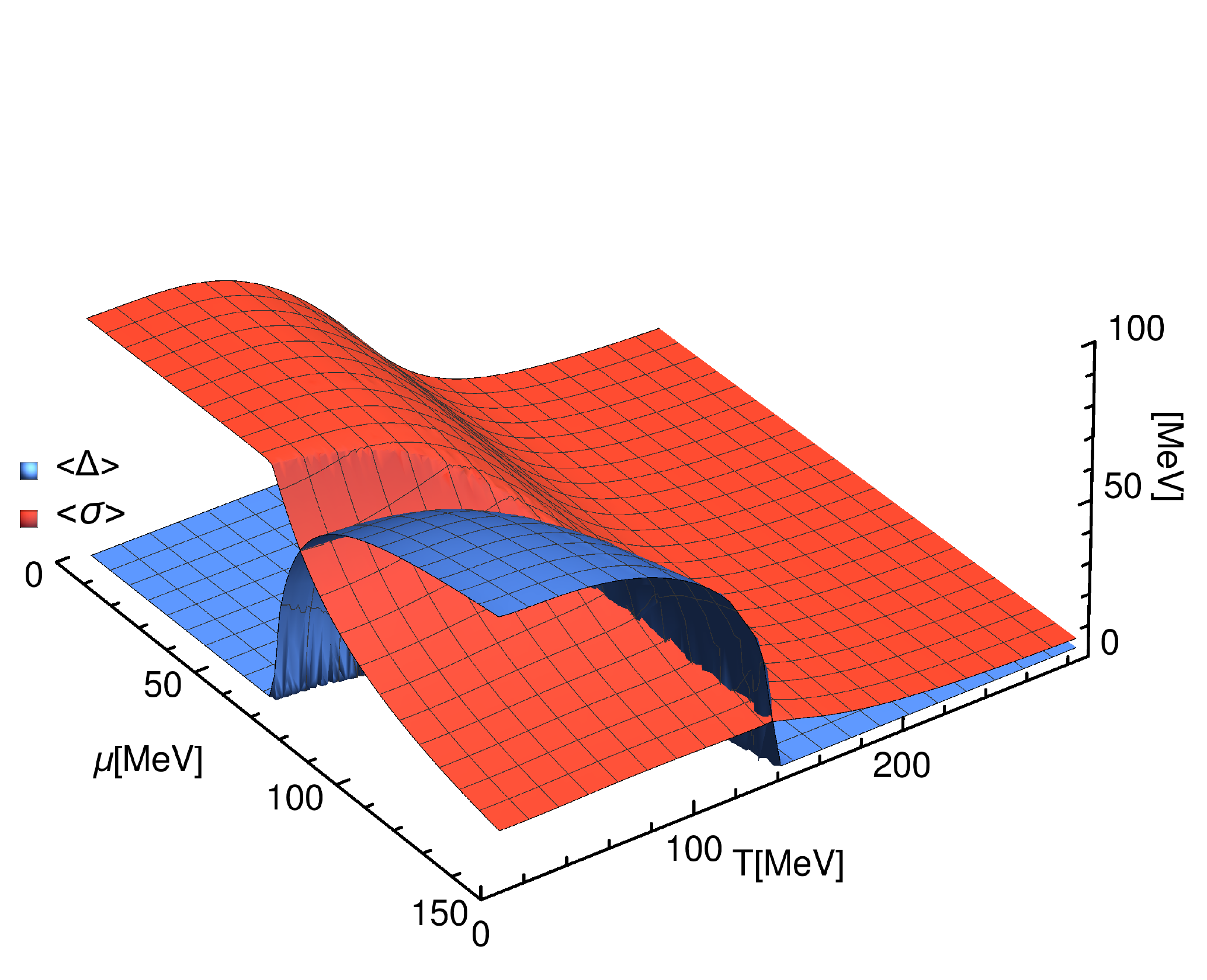}
\includegraphics[scale=0.5, trim=-1.2cm 0 0 0]{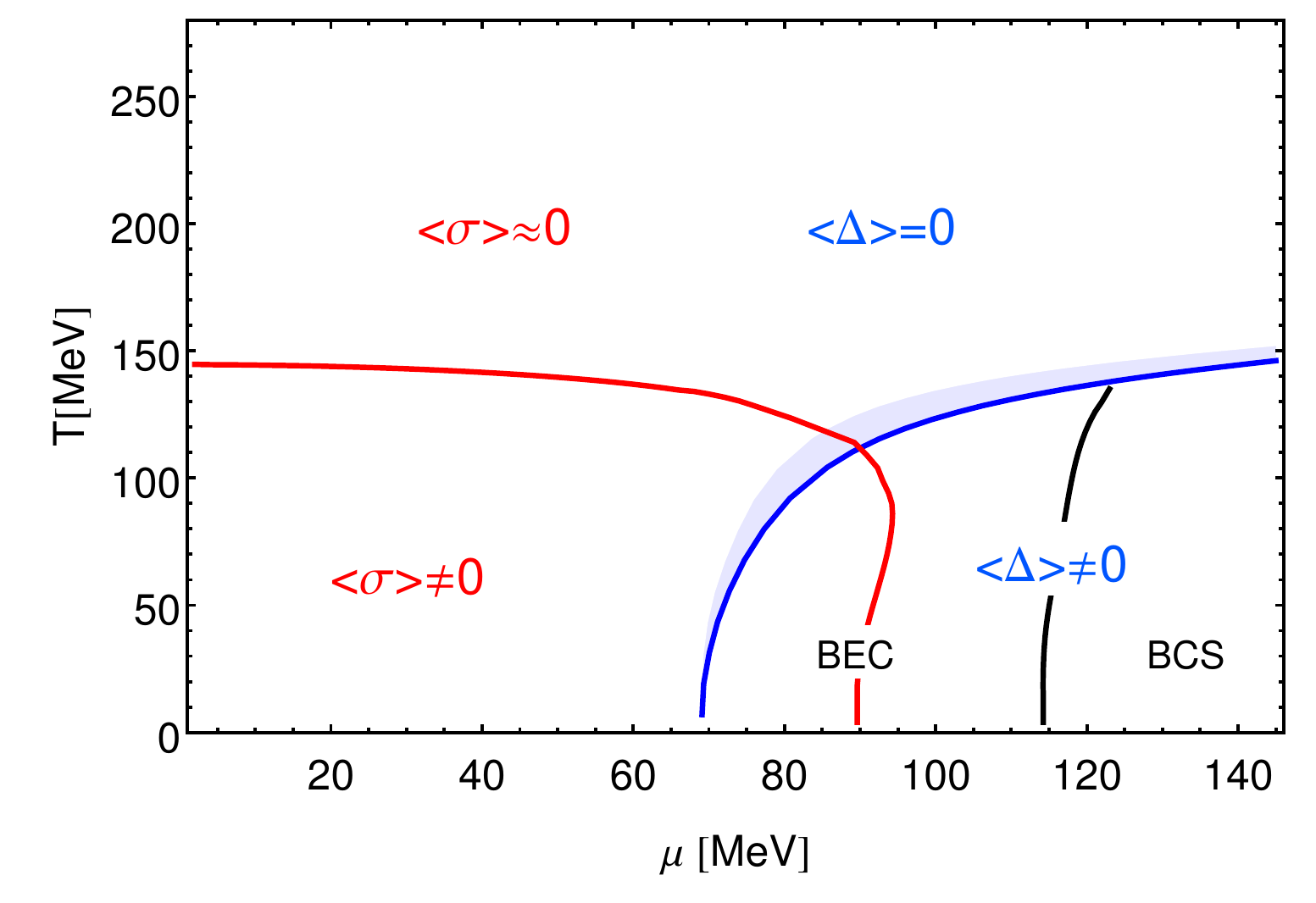}
\caption{The phase diagram of QC${}_2$D in the 2d Taylor${}^\prime$
  truncation. Right panel: The red line marks the chiral phase
  boundary which is defined by the maximum of the chiral
  susceptibility, see Ref.~\cite{Pawlowski:2014zaa}. The blue line is
  the boundary of the diquark/superfluid phase defined by
  $\lim_{k\rightarrow 0} (m_{\sp \D}-\m_{\sp\D})=0$. The blue shaded
  area marks the pre-condensation phase. The black line of the BEC-BCS
  crossover is where $h_{\sp \f}\langle \s \rangle \!=\! \mu$.}
\label{fig:QC2DPhasenDiagramm}
    \end{figure*}
%%%%%%%%%%

\subsection{Pre-condensation}\label{sec:pre-condensation}

The RG flows of the order parameters of our model close to the BEC
transition exhibit a peculiar behaviour known as pre-condensation. In
Fig.~\ref{fig:Precondflow} we show the flows of the diquark mass and
condensate as well as the chiral condensate at $\mu \!=\!
100\,\text{MeV}$ for various temperatures in the vicinity of the BEC
transition. We observe that at temperatures slightly above $T_c$ the
flowing diquark mass vanishes at an intermediate scale. Below this
scale a non-vanishing diquark condensate appears and vanishes again
before all fluctuations are integrated out towards $k=0$, i.e., the
system ends up in the chirally symmetric phase, see the left panel
Fig.~\ref{fig:Precondflow}. We refer to this behaviour as
pre-condensation. This phenomenon is an analogue to the one known from
the ferromagnetic transition close to the Curie temperature. It has
also been observed within FRG-applications, e.g., in ultra cold atoms
\cite{Boettcher:2012cm}. Here, we show for the first time that
pre-condensation is also a phenomenon in QC${}_2$D. The occurrence of
a diquark condensate at intermediate scales means that small domains
of non-vanishing diquark condensates develop which eventually average
out upon considering larger volumes, see Fig.~\ref{fig:domains}. The
left panel of Fig.~\ref{fig:Precondflow} shows that the regime of
intermediate diquark condensation grows with decreasing
temperature. Hence, these domains become larger until a single domain
fills the entire volume below the critical temperature, cf
Fig.~\ref{fig:domains}.

Pre-condensation also manifests itself in the chiral condensate. The
corresponding flow is shown in the right panel of
Fig. \ref{fig:Precondflow}. The finite range of scales where
intermediate diquark condensation occurs, gives rise to a plateau in
the chiral condensate in the same range. This can be related to a
qualitative change in the bosonic contributions to the flow of the
chiral condensate in the presence of an intermediate diquark
condensate and the concomitant Goldstone fluctuations.

%%%%%%%%%%%%%%%%%%%%%%%%%%%%%%%%%%%
%%%%%%%%%%%%%%%%%%%%%%%%%%%%%%%%%%%

%%%
%%%%%%%%%%
 \begin{figure*}[t]
\includegraphics[width=\columnwidth]{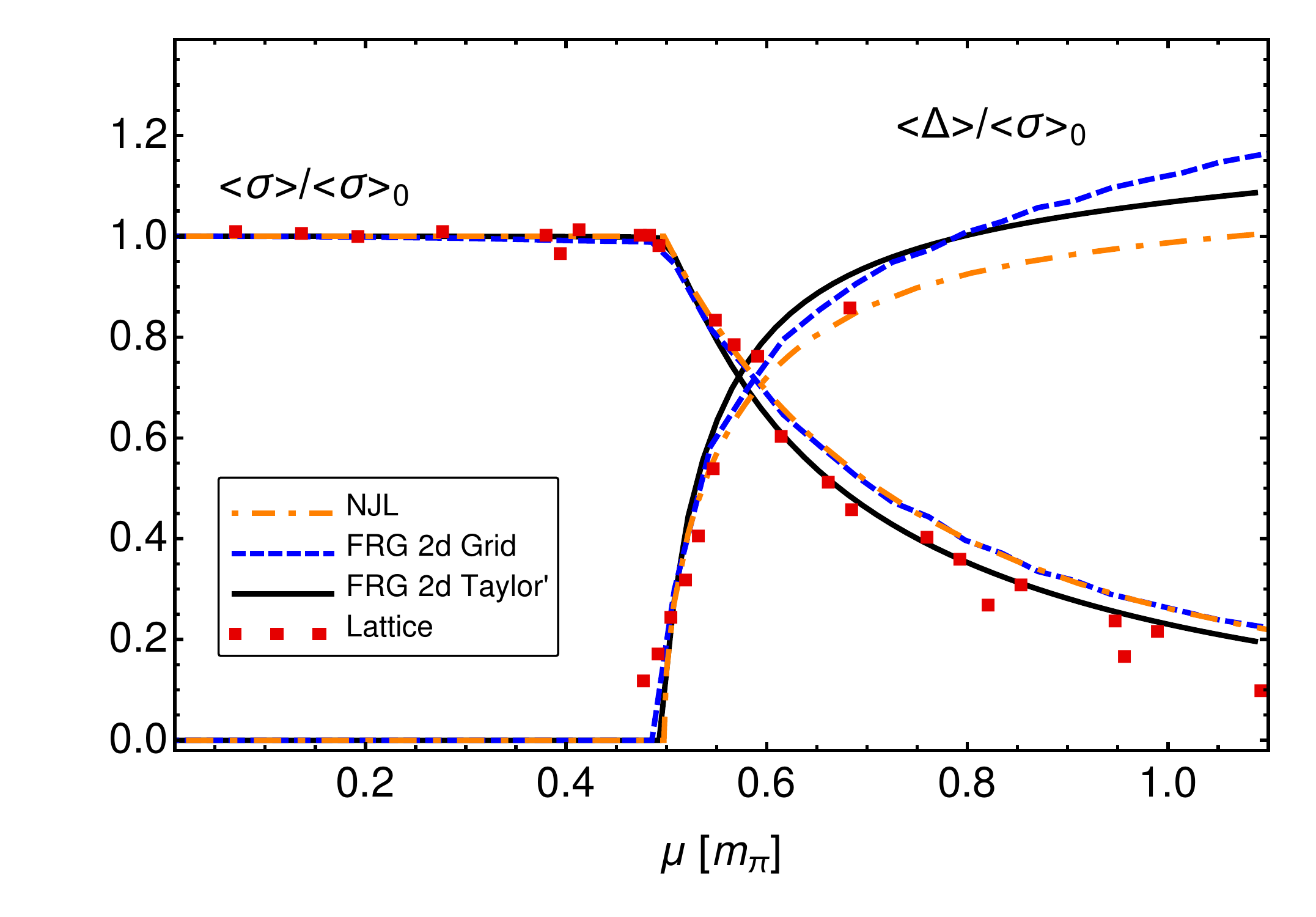}
%\hspace{0.6cm}
\includegraphics[width=\columnwidth]{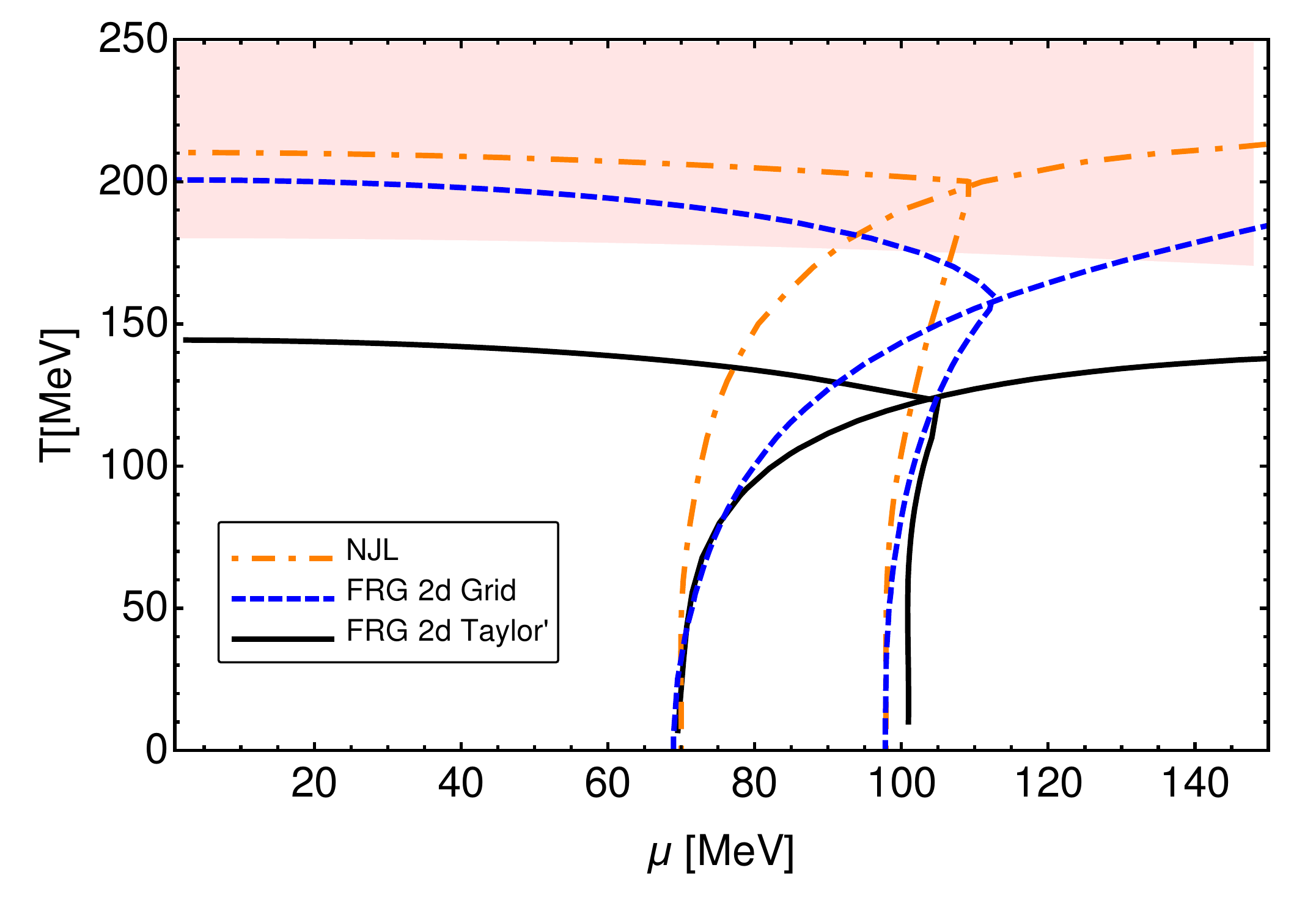}
\caption{Meson and baryon condensates at vanishing temperature (left
  panel) as well as the phase diagram (right panel) for various
  truncations and methods. The ``FRG 2d Grid" results are taken from
  \cite{Strodthoff:2011tz} and the NJL results from
  \cite{Ratti:2004ra}. The lattice data is from
  \cite{Hands:2000ei}. Chiral phase boundaries are computed via the
  half value of the order parameters and the red shaded area indicates
  the region where the effective models become invalid.}
\label{fig:QC2DPhasenAll}
    \end{figure*}
%%%%%%%%%%%%%

\subsection{The QC${}_2$D Phase Diagram}\label{sec:phasediag}

The phase diagram of two-color QCD within our full truncation is
presented in Fig. \ref{fig:QC2DPhasenDiagramm}, showing the chiral
phase transition and the occurrence of the diquark condensate as a
function of chemical potential and temperature. At small chemical
potential and small temperatures the pronounced quark fluctuations
drive chiral symmetry breaking leading to a non-vanishing chiral
condensate. Here, an increase of the temperature causes a suppression
of quark fluctuations and therefore induces a smooth crossover to the
chirally symmetric regime. Bosonic fluctuations additionally
contribute to a flattening of the crossover.

At vanishing temperature, we obtain the onset of diquark condensation
at a chemical potential of $\m=m_\pi/2$ via a second order phase
transition. For diquark chemical potentials $\m_{\sp\D}=2\m$ exceeding
their excitation gap or mass, the system is populated with diquarks,
forming the condensate at small temperatures.  The diquark condensate
triggers a decay of the chiral condensate because the condensation of
quarks into quark-quark states is preferred over quark-antiquark
pairing. Due to the small mass of the baryonic bound states, the
chiral condensate starts decaying smoothly at much smaller chemical
potentials as compared to QCD and there is no trace of a critical
endpoint.  Furthermore, a BCS pairing of quarks with opposite momenta
on the Fermi surface occurs when the chemical potential exceeds the
quark mass, cf. Sec.~\ref{ssec:FleqQMD}.

In the left panel of Fig. \ref{fig:QC2DPhasenAll}, we explicitly
compare the results for the condensates at temperature $T=0$ obtained
in various works, namely lattice calculations \cite{Hands:2000ei}, the
NJL model \cite{Ratti:2004ra}, the tree-level linear sigma model
\cite{Andersen:2010vu}, different FRG approaches
\cite{Strodthoff:2011tz} and this work. All methods agree
qualitatively, however, at larger chemical potential sizeable
quantitative corrections can be observed.  For the different FRG
approaches, these differences can be related to the additional
fluctuation effects induced by the running wave function
renormalisations and the running Yukawa couplings which have been
omitted in the LPA calculation in \cite{Strodthoff:2011tz}. We further
note that the diquark condensate obtained from the purely fermionic
NJL model agrees well with the lattice results where bosons are heavy.

In the right panel of Fig.~\ref{fig:QC2DPhasenAll}, we compare the
corresponding finite temperature phase diagrams showing results from
the NJL model and the various FRG approaches. Here, the chiral
crossover is defined by the half value of the order parameter. The
considerable quantitative effects within the FRG approach, induced by
taking into account the running of Yukawa couplings and wave function
renormalisations are clearly exposed by comparing 2d~Taylor$^\prime$´
and 2d~Taylor results for the critical temperature at vanishing
density. While we find $T_c\approx 150$ MeV for 2d~Taylor$^\prime$´,
see Fig.~\ref{fig:QC2DPhasenAll}, we have $T_c\approx 210$ MeV, see
Fig.~\ref{fig:allfrgpds} for the 2d~Taylor approximation with the same
initial condition. Indeed the latter critical temperature is found for
all LPA approximations with different initial conditions, see
Fig.~\ref{fig:allfrgpds}. This underlines the quantitative importance
of the wave function renormalisation and the Yukawa coupling. Hence,
non-classical dispersion relations as well as quantum corrections to
the quark-boson interactions play a crucial role for an accurate
description of the phase structure of QC${}_2$D. To corroborate this
statement, we demonstrate in App.~\ref{app:conv} that this large
effect cannot be ascribed to different initial conditions.

\begin{figure*}[t]
    \includegraphics[scale=0.52, trim=1cm 0 0 0]{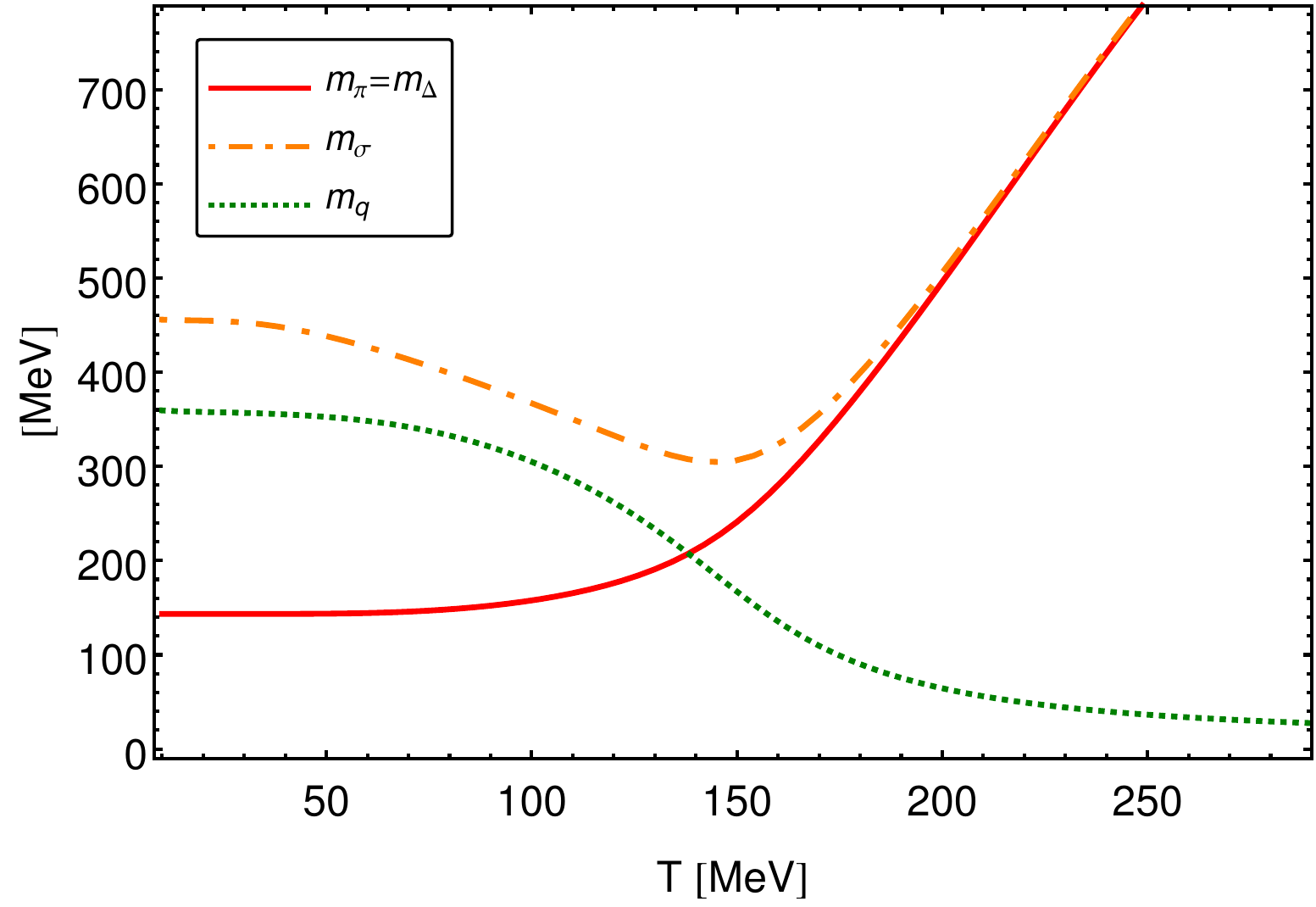}
\includegraphics[scale=0.52, trim=-2cm 0 0 0]{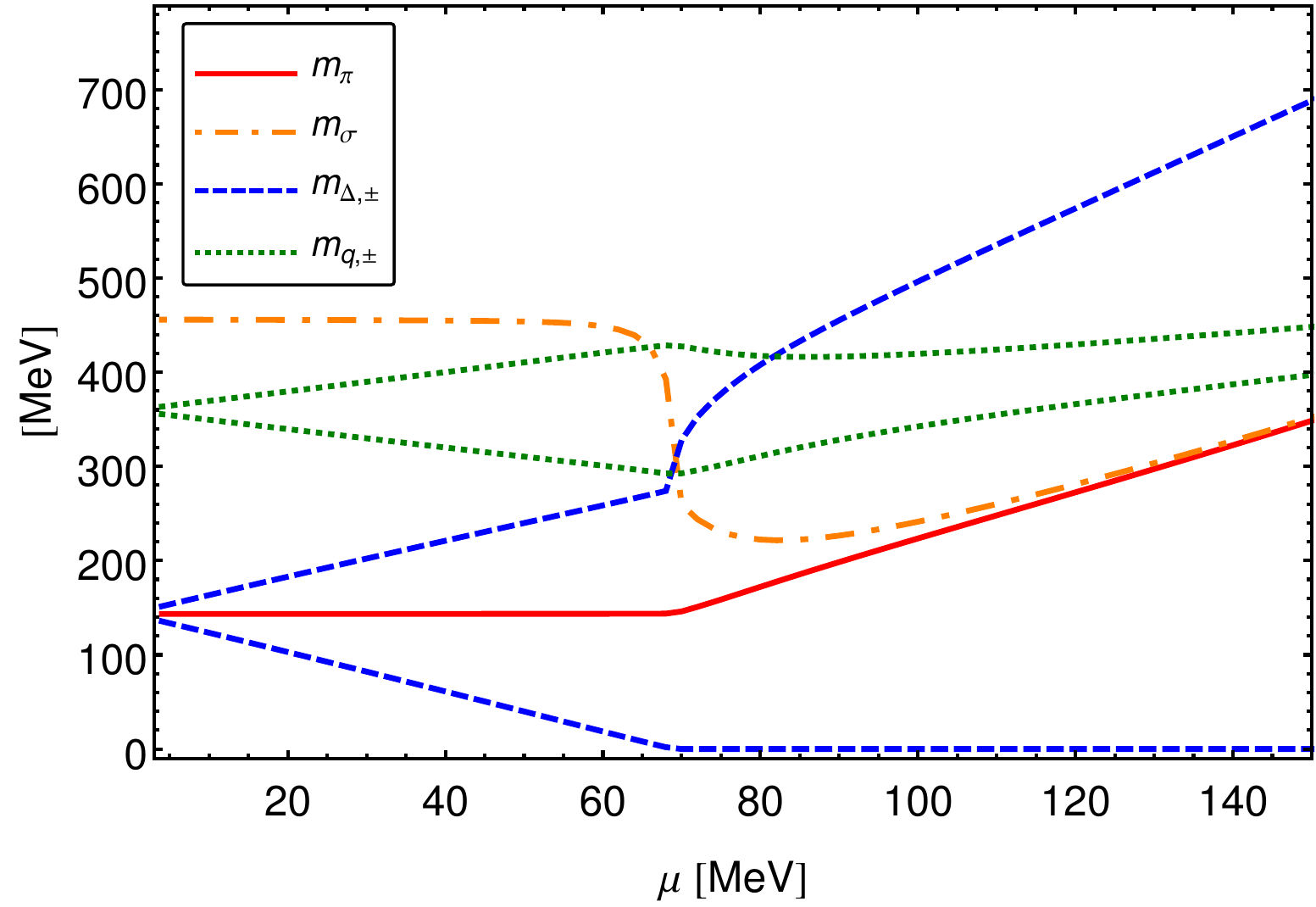}
\caption{Mass spectrum of two-color QCD within the 2d
  Taylor${}^\prime$ truncation. The chiral crossover and the phase
  transition to the BEC phase are clearly visible from the behaviour
  of the masses. In the left panel ($\mu\approx 0$) the quark mass
  goes from being small to being large with decreasing temperature,
  while the pions and diquarks become the pseudo-Goldstone modes
  splitting up from the sigma. In the right panel ($T\approx 0$) we
  have a phase transition at $\mu=\frac{m_{\p}}{2}$. Even though
  chiral symmetry is restored at asymptotic $\m$, the quarks do not
  become massless because they are coupled to the diquark condensate.}
\label{fig:QC2Dmass}
\end{figure*}

The NJL result has the highest critical temperatures for both, the
chiral crossover and the superfluid phase transitions. In the FRG
computation within the LPA the phase boundaries are lowered by about
10\% through the inclusion of symmetry-restoring bosonic
fluctuations. The additional effects considered in the present work
lead to a further decrease of the phase boundary of about 30\%,
cf. Fig.~\ref{fig:QC2DPhasenAll}, see App.~\ref{app:conv} for details.
At high temperatures, the cutoff scale $\Lambda$ constrains the
applicability of the effective low-energy description presented here,
as indicated by the red-shaded area in
Fig.~\ref{fig:QC2DPhasenAll}. Here, the initial conditions become
effectively temperature and density dependent, see
App.~\ref{app:medcut} for details.

%%%%%%%%%%%%%%%%%%%%%%%%%%%%%%%%%%%
%%%%%%%%%%%%%%%%%%%%%%%%%%%%%%%%%%%

\subsection{Mass Spectrum}

Now, we discuss the curvature masses of mesons and diquarks for the
normal as well as for the BEC phase in the meson sector
\begin{align}
  &\text{normal:}& m_{\p}=m_{\sp \f}\,, &\quad m_{\s }=
  \sqrt{m^2_{\sp\phi }+2\k_{\sp\phi }\l_{2,0}}\,,\nonumber\\
  &\text{BEC:} & m_{\p}=m_{\sp \f}\,, &\quad
  m_{\tilde\s}=\sqrt{m_{1}^{2}-m_{2}^{2}} \,, \label{eq:BECmasses1}
\end{align}
and in the baryon sector
\begin{align}
  &\text{normal:}& m_{{\sp \D},+}=m_{\sp \D}+2\mu\,, &\quad m_{{\sp
      \D},-}
  =m_{\sp \D}-2\mu\,,\nonumber\\
  &\text{BEC:} & m_{\tilde{\sp \D}_{1}}=\sqrt{m_{1}^{2}+m_{2}^{2}}
  \,,&\quad m_{{\sp \D}_{2}}=0\;. \label{eq:BECmasses2}
\end{align}
Here, we have defined the mass parameters
\begin{align}
  m_{1}^{2}&=\frac{m_{\s}^{2}}{2}+8\mu^{2}+\k_{\sp \D}\l_{0,2}\;,\nonumber\\
  m_{2}^{2}&=\sqrt{m_{1}^{4}-2\k_{\sp
      \D}\left(m_{\s}^{2}\l_{0,2}-2\k_{\sp
        \f}\l_{1,1}^{2}\right)-16m_{\s}^{2}\mu^{2}}\;, \nonumber
\end{align}
and $m_{\s}$ is the expression for the sigma mass in the normal
phase. The sigma state and one of the diquark states are mixtures of
the original states. Further, ${\D}_{2}$ is the Goldstone mode
associated with the breaking of $U(1)_{B}$.

In Fig.~\ref{fig:QC2Dmass}, we present the temperature and chemical
potential dependence of the masses. At vanishing chemical potential,
see left panel of Fig.~\ref{fig:QC2Dmass}, the pion and the diquark
masses are degenerate. For large temperatures, where chiral symmetry
is restored, they join with the sigma mass and all bound states
decouple from the system.  For vanishing temperature, see right panel
of Fig.~\ref{fig:QC2Dmass}, the diquark starts condensing at
$\mu_{c}=\frac{m_{\p}}{2}$ and the behaviour of the masses changes
accordingly.  At large chemical potential chiral symmetry is restored
and $m_{\tilde \s}\ra m_{\sp \f}$. Then, the sigma and the pions are
degenerate. Accordingly, the corresponding (anti-)quark masses are
given by the relation \be m_{q,\pm}=\sqrt{(\hat \s \pm \mu)^{2}+\hat
  \D^{2}}\;, \ee and again chiral symmetry is restored at large $\mu$.

%%%%%%%%%%%%%%%%%%%%%%%%%%%%%%%%%%%
%%%%%%%%%%%%%%%%%%%%%%%%%%%%%%%%%%%

\section{Conclusions}\label{sec:conc}

In this work, we add important aspects to the understanding of the
high density regime of QCD by investigating a modification of QCD with
two colors. The absence of a fermion sign problem in this model has
drawn the interest of the lattice community for over a decade
now. Here, we refine an alternative non-perturbative approach based on
the functional renormalisation group.

More specifically, we employ an effective low-energy model for
QC${}_2$D where the gluon degrees of freedom have been integrated
out. This amounts for a description of the theory in terms of quarks,
mesons and diquarks. The diquarks are color neutral objects and
constitute the baryons of the theory. Therefore, we have a playground
at hand to study baryonic degrees of freedom in a simplified bosonic
description as well as a relativistic BEC-BCS crossover.  The
FRG-approach allows to interpolate between the microscopic and the
macroscopic regime of the theory, by integrating out all thermal and
quantum fluctuations. It generates all correlations that are allowed
by the underlying symmetries. By examining the flow equation of the
effective potential, we study the interplay of bosonic and fermionic
degrees of freedom for different temperatures and chemical potential.
The chiral condensate goes through a crossover towards large
temperatures, where chiral symmetry is restored. The diquark
condensate sets in via a second order phase transition at large
chemical potential and small temperatures.

For the first time in the context of two-color QCD, we trace the
momentum scale dependence of the diquark condensate and exhibit the
phenomenon of pre-condensation: Small domains of non-vanishing diquark
condensates appear at intermediate RG scales, however, no finite
condensate persists as all fluctuations are integrated out. Therefore,
the pre-condensation phase is a precursor of the BEC phase.  Further,
we find that the phase boundaries in our full truncation are
significantly corrected upon inclusion of dynamical quark-hadron
interactions and non-classical dispersions. We conclude that an
inclusion of these effects is an important ingredient for a reliable
determination of the phase boundaries in the phase diagram of
QC${}_2$D. Most strikingly, the chiral phase transition does not have
a critical endpoint in the presence of diquark fluctuations.

Eventually, as an essential aspect of this work, we thoroughly studied
the Silver Blaze property in the FRG framework. We exhibited that at
vanishing temperature it amounts to a shift of the frequencies by the
chemical potential with the appropriate charges. In case this property
holds true at the initial scale, it is preserved by the flow equation,
owing to its one-loop structure. However, this requires a full
momentum resolution of all quantities, which are directly sensitive to
the chemical potential.  The diquark sector is strongly affected by
this complication, if its parameters are not identified with the meson
sector.  Importantly, running wave function renormalisations
considerably reduce the discrepancy between the onset chemical
potential and the pion mass in the two-color case.

A natural next step of our work is to incorporate a phenomenological
Polyakov-loop potential to investigate confinement in two-color QCD,
cf. \cite{Strodthoff:2013cua} for a LPA study.  Further, a dynamical
connection of the quark-gluon regime with the low-energy effective
regime should be established, cf. Refs. \cite{Braun:2014ata,
  Rennecke:2015vm, Mitter:2014wpa}.

%%%%%%%%%%%%%%%%%%%%%%%%%%%%%%%%%%%
%%%%%%%%%%%%%%%%%%%%%%%%%%%%%%%%%%%

{\it Acknowledgments -} We thank Nils Strodthoff, B.-J. Schaefer and
L. von Smekal for discussions. This work is supported by the Helmholtz
Alliance HA216/EMMI, the grant ERC-AdG-290623, the FWF grant
P24780-N27, and the BMBF grant 05P15VHFC1.

%%%%%%%%%%%%%%%%%%%%%%%%%%%%%%%%%%%
%%%%%%%%%%%%%%%%%%%%%%%%%%%%%%%%%%%

\begin{appendix}\label{app}

%%%%%%%%%%%%%%%%%%%%%%%%%%%%%

\section{Truncation Effects}

%%%%
%%%%%%%%%%%%%%
\begin{figure}[b]
  \includegraphics[scale=0.65, trim=1cm 0.8cm 0 0]{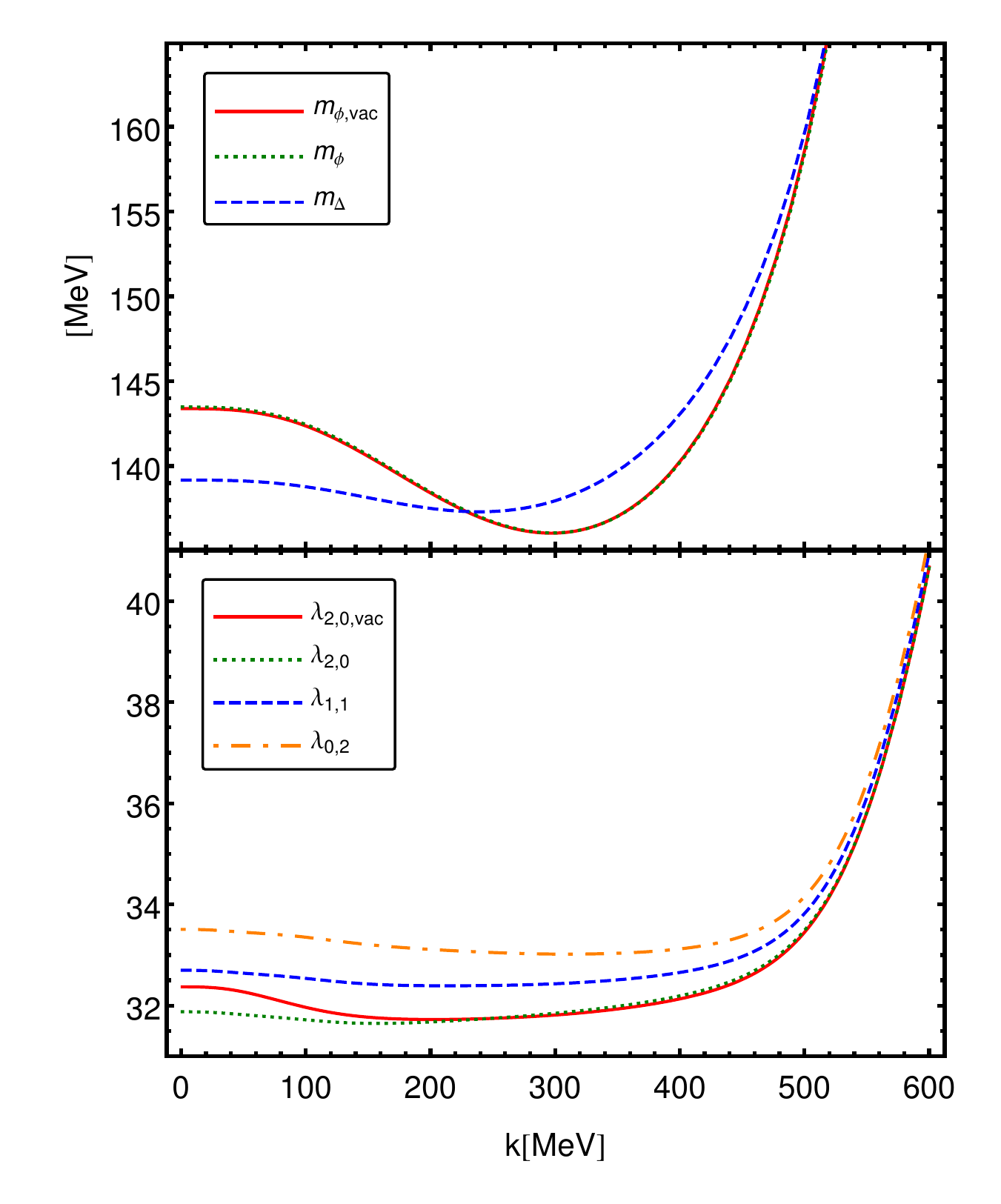}
  \caption{Flow of the mass parameters and four-point couplings. The
    red lines represent the vacuum flow, i.e. at $T\!=\!\mu\!=\!0$,
    the other lines are at $\m=0.4\, m_\pi$.}
   \label{fig:flowCoupl}
\end{figure}
%%%%%%%%%%%%%%
%%%%

In this section we study to some extent the reliability of our
truncation by analysing the violation of Silver-Blaze, the convergence
of our expansion, the effects of different parts of our truncation and
in-medium cutoff effects.

\subsection{Silver-Blaze Violation}\label{app:sbv}

In Fig.~\ref{fig:flowCoupl} we show the flows of the relevant and
marginal hadronic couplings for $T\!=\!0$ at two choices of the
chemical potential, $\m \!=\! 0$ (vacuum) and $\m \!=\! 0.4\,
m_\pi$. Due to the $SO(6)$ symmetry between meson and diquarks in the
vacuum, all coefficients $\lambda_{n,m}$ of a given order $N$ with
$n\!+\!m \!=\!  N$ are identical. They are represented by the solid
red lines in Fig.~\ref{fig:flowCoupl}. According to the Silver Blaze
property, all $n$-point functions have to be independent of the
chemical potential at vanishing temperature and for $\mu<\mu_c$. This
implies that all couplings $\lambda_{n,m}$ of a given order $N$ with
$n\!+\!m \!=\! N$ also have to be identical in this regime, i.e. also
for $\m \!=\! 0.4\, m_\pi$. The deviations of this expectation, see
Fig.~\ref{fig:flowCoupl}, therefore show the extent of the violation
of the Silver Blaze property. The higher the order of the diquark
fields (i.e. the second index) the sooner the flow starts to separate
from the vacuum flow. This also couples back to the purely mesonic
couplings. It is interesting that $\l_{2,0}$ starts to deviate from
the vacuum flow at around $k\approx 2\mu$. Nevertheless, even though
higher order terms have a large deviation, they do not couple back
very strongly to the physically important quantities: $m_{\sp
  \D}=\sqrt{\l_{0,1}}$ has only a deviation of about 5~MeV.

Following our discussions in Sec.~\ref{sec:SilvBl} and
\ref{ssec:FleqQMD}, this violation is well understood. Owing to our
momentum independent expansion scheme, we generate explicitly
$\mu$-dependent couplings. Then the chemical potential
enters the flow of the effective potential through curvature
contributions in diquark direction. Therefore, the higher the
order of the diquark fields in a given $n$-point function, the larger
the violation of Silver Blaze. However, our results presented in the
previous section show only a minor violation implying that the
back-coupling of the violation in the $n$-point functions to the
physical observables is small.

\subsection{Convergence of the Expansion}\label{app:conv}

%%%
%%%%%%%%%%%
 \begin{figure}[!t]
\includegraphics[width=.96\columnwidth]{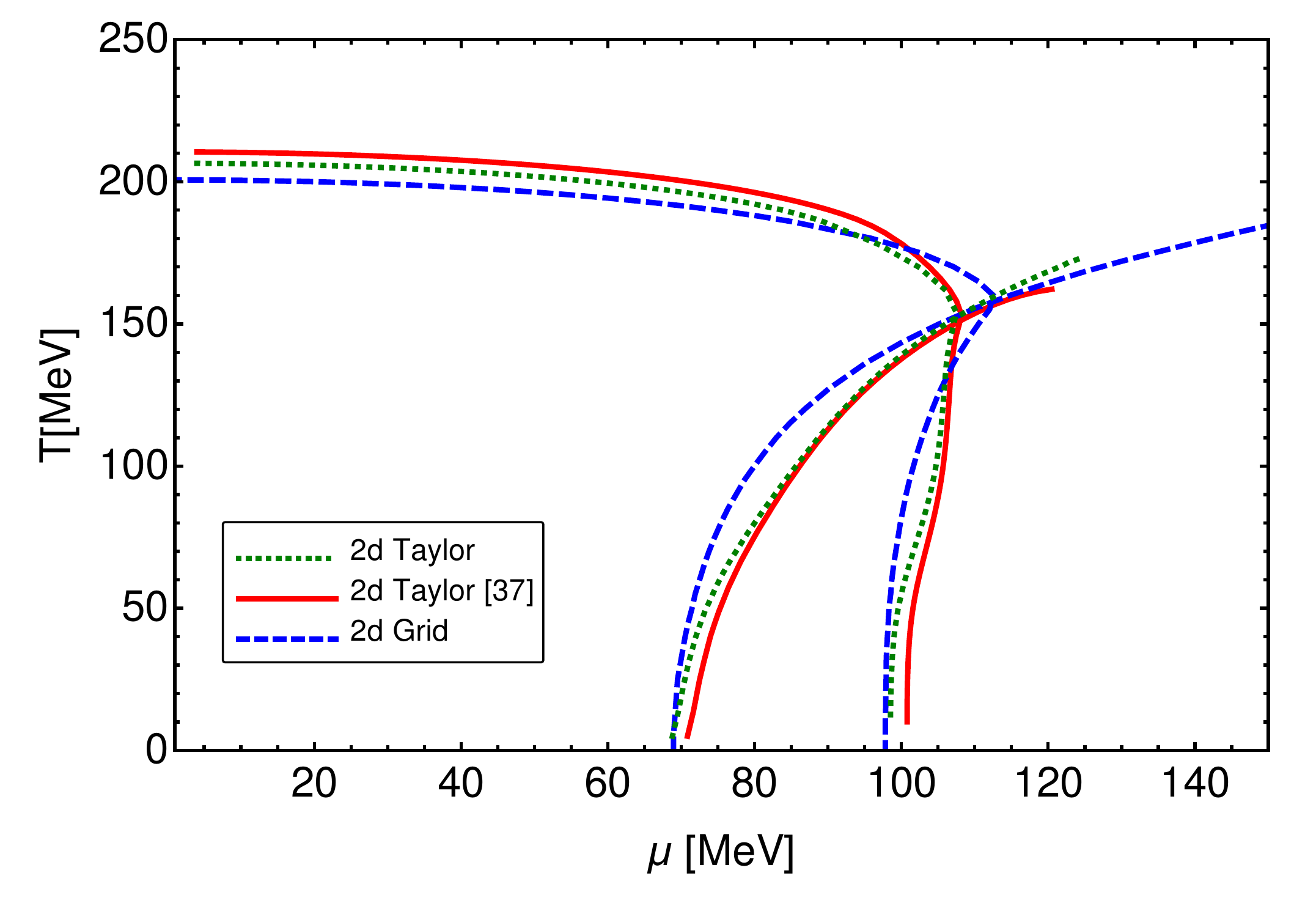}
\caption{The phase boundaries from the grid method
  \cite{Strodthoff:2011tz} and the Taylor method with different
  initial conditions. The initial conditions of ``2d Taylor [37]" are
  the same as in the reference. Hence, A comparison between solid red
  line and the dashed blue line can be used to estimate how well the
  Taylor expansion is converged. The initial conditions of ``2d
  Taylor" feature comparably weakly interacting heavy mesons at the
  initial scale. A comparison between the dotted green line and the
  solid red line can therefore be used to estimate the effect of
  different initial conditions on the phase boundaries. The initial
  conditions are given in Tab.~\ref{tab:initial} and are discussed in
  Sec.~\ref{sec:initconds}.}
\label{fig:QC2DPhasenDA}
    \end{figure}
%%%%%%%%%%%
%%%

    Here we study the convergence of our Taylor expansion.  To measure
    its quantitative error we compare the results in the LPA to the
    results obtained on a two-dimensional grid with the same
    truncation in \cite{Strodthoff:2011tz}. This is shown in
    Fig.~\ref{fig:QC2DPhasenDA}. The deviation between the
    two-dimensional Taylor expansion (solid red line) and the grid
    solution (dashed blue line) can be traced back to the lack of
    convergence in the LPA part of our truncation, cf. also
    App.~\ref{app:1dt}. However, we have shown in
    Sec.~\ref{sec:phasediag} and in particular the right panel of
    Fig.~\ref{fig:QC2DPhasenAll} that the error of neglecting effects
    beyond LPA is about 30\%. Hence, regarding the systematic
    quantitative improvements within the present work, the error
    related to the expansion of the effective potential is only
    minor. We will discuss the effects that go beyond the LPA in the
    next section.

    Interestingly, we also see in Fig.~\ref{fig:QC2DPhasenDA} that
    there is only a small difference between the phase boundaries
    obtained with the initial conditions shown in the second and third
    row of Tab.~\ref{tab:initial}, named ``2d Taylor" (dotted green
    line) and ``2d Taylor [37]" (solid red line) respectively. We note
    that the initial conditions ``2d Taylor" are chosen in the same
    spirit as the initial conditions for our full truncation: As
    appropriate for the chirally symmetric phase, the bosonic
    interactions are weak and bosons are heavy and therefore decoupled
    from the system. Hence, as discussed in Sec.~\ref{sec:phasediag},
    the large difference between the phase diagram from the LPA in
    \cite{Strodthoff:2011tz} and from our full truncation in
    Fig~\ref{fig:QC2DPhasenAll} can be attributed to the highly
    relevant fluctuation effects that induce the running of the Yukawa
    couplings and the wave function renormalisations.

\subsection{Truncation Effects Beyond the LPA}\label{sec:blpa}
    
\begin{figure}[t]
\includegraphics[scale=0.5, trim=0 1.85cm 0 0, clip=true]{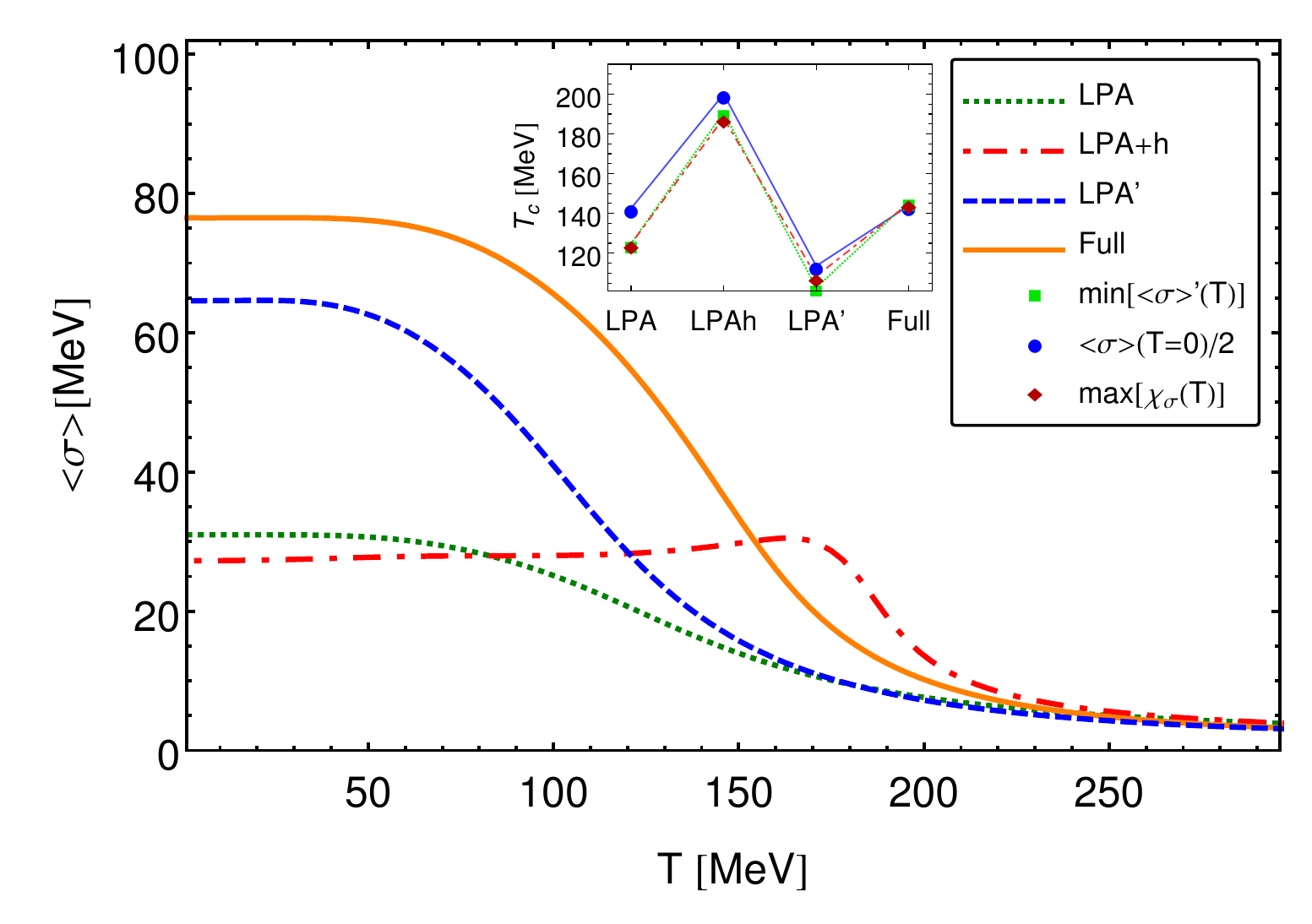}
\caption{The chiral condensate as a function of temperature and the
  transition temperatures at $\m=0$ for different truncations. The
  Taylor expansion is to the order $N=5$.}
\label{fig:Truncs}
    \end{figure}

Here, we will investigate the truncation effects beyond the LPA. In
the present case, we study the effect of a running meson and diquark
Yukawa couplings and running quark and hadron wave function
renormalisations. In Fig.~\ref{fig:Truncs} we show the chiral
condensate as well as the critical temperature for different subsets
of our full truncation~(\ref{eq:QMDaction}). Here, ``LPA$^\prime$" has
the running effective potential as well as the running quark, meson
and diquark wave function renormalisations, but a constant bare Yukawa
coupling. Note that the renormalised Yukawa couplings have a trivial
running induced by the non-vanishing anomalous dimensions. On the
other hand, ``LPA+h" takes the running of the effective potential and
the Yukawa couplings into account, but has vanishing anomalous
dimensions.

\begin{figure*}[!t]
     \includegraphics[scale=0.55, trim=1cm 0 0 0]{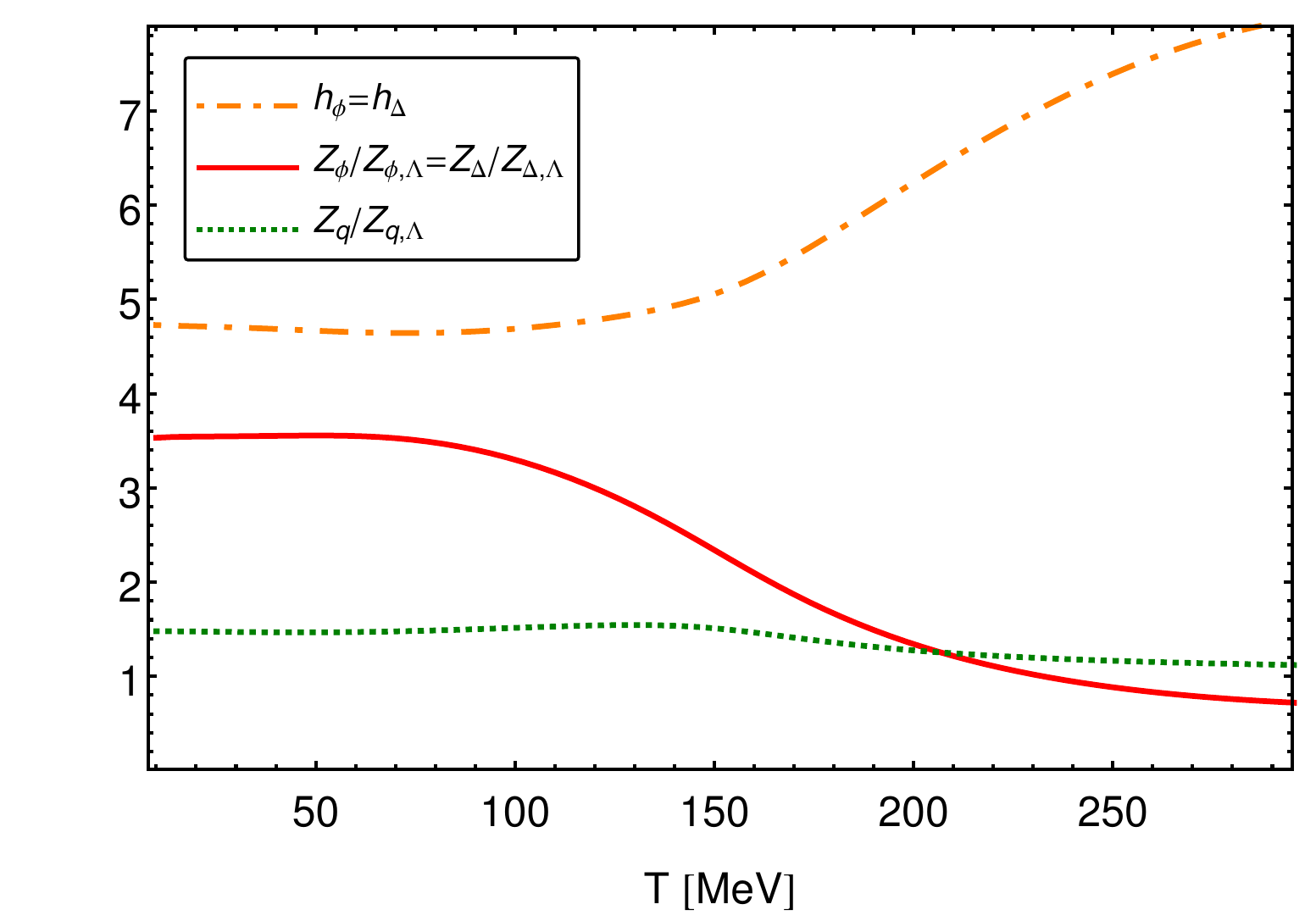}
 \includegraphics[scale=0.55, trim=-1cm 0 0 0]{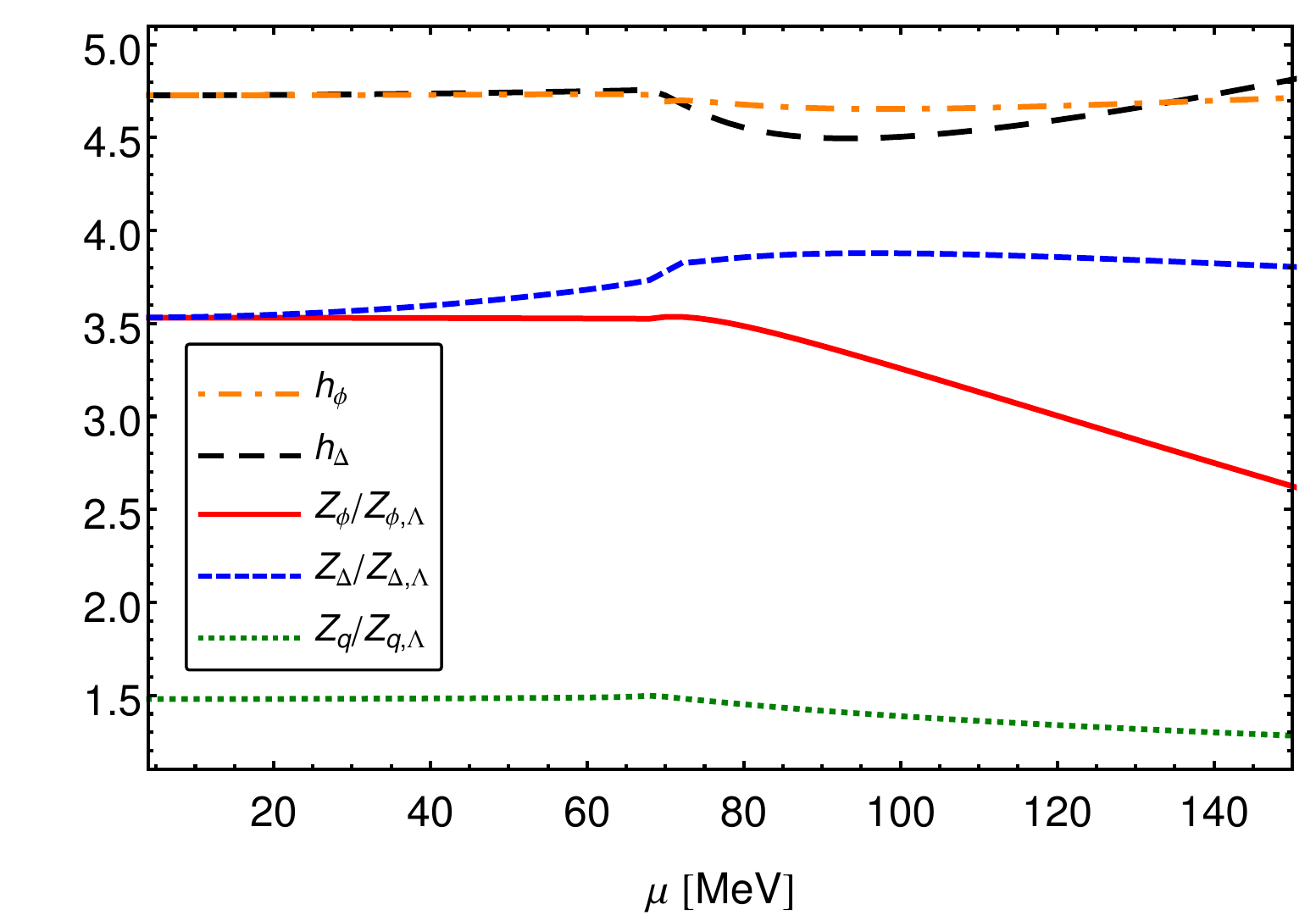}
 \caption{The Yukawa couplings and the wave function renormalisations
   as functions of $T$ at $\mu\!=\!0$ (left) as well as functions of
   $\mu$ at $T\!=\!0$ (right). The wave function renormalisations are
   normalised to their initial values in the UV. The explanations of
   each parameter can be found in the respective paragraphs.}
 \label{fig:HsUndZs}
     \end{figure*}

We see large effects on both, the chiral order parameter and the
critical temperature. In general, we conclude that for fixed initial
conditions and compared to LPA, the Yukawa coupling increases the
critical temperature, while the running wave function renormalisations
decrease it. The former effect is due to increased quark fluctuations
which, in turn, lead to a larger phase of broken chiral symmetry. The
latter is a result of the large positive bosonic anomalous dimensions
in the vicinity of the phase transition, see
e.g. Fig.~\ref{fig:HsUndZs}. Since the initial masses in the UV are
fixed for all truncations, the meson masses in LPA$^\prime$ are driven
to smaller values by the increasing wave function renormalisations
towards the IR as compared to LPA. Hence, bosonic fluctuations lead to
a larger symmetric phase in the $T\!-\!\mu$ plane. This has also been
observed in a quark-meson model for low-energy QCD in
\cite{Pawlowski:2014zaa}. At fixed initial conditions the effects of
running Yukawa couplings and wave function renormalisations almost
cancel for the critical temperature. This is in contrast to the case
where for each truncation the initial conditions where tuned to the
phenomenological IR parameters as it is shown in the right panel of
Fig.~\ref{fig:QC2DPhasenAll}. There, we see that the critical
temperatures of LPA and our full truncation differ by about 30\%.

%%
%%%%%%
%%%%%%
%%

%%%
%%%%%%%%%%%%
%%%%%%%%%%%%
%%%

     Finally, we want to take a closer look at the in-medium
     modifications of the wave function renormalisations and the
     Yukawa couplings. In Fig.~\ref{fig:HsUndZs} the
     temperature/chemical potential dependence of the wave function
     renormalisations and the Yukawa couplings is shown. We see that
     the bosonic wave function renormalisations indicate in which
     regime the mesons/diquarks become auxiliary fields. Decreasing
     bosonic wave function renormalisations, e.g. towards large
     temperatures, imply that the corresponding fields become
     auxiliary \cite{Braun:2014ata, Rennecke:2015vm}. The left figure
     shows that, while the quark wave function renormalisation does
     not change much, the meson and diquark wave function
     renormalisations decrease in the vicinity of the critical scale
     with increasing temperature and at vanishing density. This
     implies a suppression/decoupling of bosonic fluctuations in the
     chirally symmetric phase.

The right plot shows the behaviour of the wave function
renormalisations as a function of the chemical potential at vanishing
temperature. Upon entering the BEC-phase, diquark fluctuations are
enhanced, while meson fluctuations decouple.

Furthermore, despite the momentum-independent approximation we employ
in this work, we see only a very weak dependence on the chemical
potential below $\m_c \!=\! m_\p/2$ of the quantities shown in the
right plot of Fig.~\ref{fig:HsUndZs}. The largest violation of the
Silver Blaze property is visible in the diquark wave function
renormalisation. This is not surprising since this is a purely baryonic
quantity. In general, the mild dependence on $\mu$ below $\mu_c$
hints at a rather mild frequency dependence of these quantities. This
justifies the use of the lowest order derivative expansion in the
present work a posteriori.

%%%%%%%%%%%
%%%%%%%%%%%

\subsection{In-Medium Cutoff Effects}\label{app:medcut}

\begin{figure}[t]    
\includegraphics[width=.95\columnwidth]{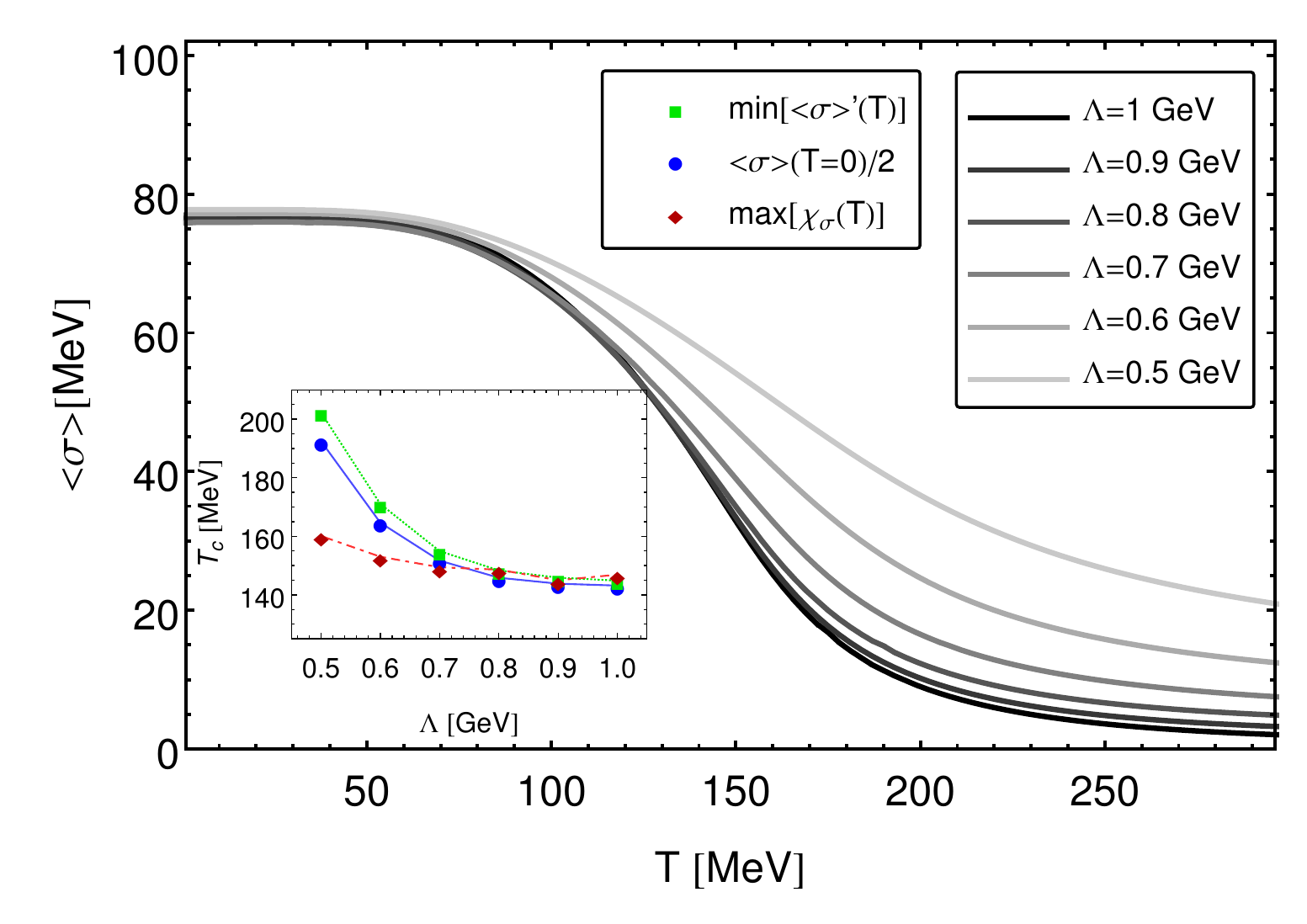}
\caption{Chiral condensate and transition temperatures at $\m=0$ for
  different cutoff scales $\Lambda$.}
\label{fig:medcut}
\end{figure}

In Fig. \ref{fig:medcut} we show the cutoff dependence of the chiral
order parameter. For this study we used the initial conditions in the
second row of Tab. \ref{tab:initial} and solved the flow equations
down (or up) to each of the $\Lambda$'s shown in the figure in the
vacuum. Note that we tuned our initial conditions at $\Lambda
\!=\!900\,\text{MeV}$. Then each time we used this as the starting
point for the finite temperature calculations. However, we always
started with a quartic potential in the UV, meaning that the higher
order coupling are set to zero at every $\Lambda$. This is why we see
small deviations at $T=0$, which can be regarded as a subsidiary probe
for the convergence of the Taylor expansion. In any case, we clearly
see rising deviations at large temperatures with the lowering of the
cutoff. If the microscopic action feels the in-medium effects, it is
not really microscopic. In this regime, the scale of thermal
fluctuations exceeds the initial scale. In particular the crossover
temperature should be independent of the cutoff. In the inset we
observe that for cutoffs $\Lambda> 800$ MeV it starts to converge.

The red shaded area in Fig. \ref{fig:QC2DPhasenAll} marks the region
where
\be \left|\frac{\dot\G_{\Lambda,T,\m}-\dot\G_{\Lambda,0,0}}{
    \dot\G_{\Lambda,0,0}}\right|>0.1\,. \label{eq:validaty} 
\ee 
This measures the difference between the initial flow of the effective
action in the medium and in the vacuum. If this quantity is large, it
means that the microscopic action is considerably influenced by
in-medium effects. In order to guarantee cutoff-independence of our
results, the UV-cutoff has to be chosen large, i.e. $\Lambda \gtrsim
800\,\text{MeV}$. On the other hand, the initial scale should not be
larger than the scale of validity of the model. For our effective low
energy models we must remain below the scale were gluon degrees of
freedom are relevant. Thus, $\Lambda$ is constrained from both sides.

%%%%%%%%%%%%%%%%%%%%%%%%%%%%%%%%%%%
%%%%%%%%%%%%%%%%%%%%%%%%%%%%%%%%%%%

\section{Propagators and Vertices} \label{app:prop}

The propagator is defined by taking two functional derivatives of the
effective action (\ref{eq:QMDaction}) with respect to the fields,
adding the regulator and then evaluating at the expansion point,
before taking the inverse. We employ a 3d flat regulator. Hence,
only the space-like momenta are regularised, $\vec p \rightarrow \vec
p_r$, with
\begin{align}
 \vec{p}_r&=
 \left\lbrace
\begin{array}{cc}
  \vec{p} \sqrt{1+r_{\sp B}(\vec{p}^{\,2}/k^2)}  & \text{for bosons,}\\[2ex]
  \vec{p}\,(1+r_{\sp F}(\vec{p}^{\,2}/k^2)) & \text{for fermions.} 
  \label{eq:regularizedP}
\end{array}
 \right. 
\end{align}
Therefore we will omit the regulator terms in this section, while for
the flow equations the momenta can simply be replaced by the
regularised ones. The time-like components of the momenta are replaced
by the Matsubara frequencies $\omega_n=2n\pi T$ for bosons and
$\nu_n=(2n+1)\pi T$ for fermions.

\subsection{Bosons}\label{app:BOSONpropagator}
\vspace{-3.mm}

Here we will explicitly display the boson propagator for the real
representation $\vf=(\vec \p,\s,\D_{1},\D_{2}) $ as well as the
complex representation $\bar\vf=(\vec \p,\s,\D,\D^{*}) $. The relation
is given by $\D=(\D_{1}+i \D_{2})/2$. The expansion points are given
by $\vf_{0}=(\vec 0,\sqrt{2\r_{\sp\f}/Z_{\sp\f} },
\sqrt{2\r_{\sp\D}/Z_{\sp\D}},0) $ and $\bar\vf_{0}=(\vec
0,\sqrt{2\r_{\sp\f}/Z_{\sp\f} }, \sqrt{\r_{\sp\D}/Z_{\sp\D}},
\sqrt{\r_{\sp\D}/Z_{\sp\D}}) $, where we leave the $\r$'s as
variables. The flow equation preserves the symmetries of the action,
hence we can always rotate any other choice of $\varphi_0$ and $\bar{
  \varphi_0}$ with the corresponding symmetry group, which, at finite
chemical potential, is $O(4)\times O(2)$ for the real representation
and $O(4)\times U(1)$ for the complex representation. Since the
observable condensate is defined as $\langle|\D|\rangle$, which is
always real, we can choose it to be in real part of the diquark
field. Consequently there is a $\r_{\sp\D}$ in both diquark fields in
the complex representation. In this case, the Goldstone mode is in the
phase of the field.

The boson propagator in the real representation it is given by
\bea
G_{{\varphi}} = \left(
    \begin{array}{cccc}
      \frac{ G_\p}{Z_{\scriptscriptstyle \phi } }\,\mathds{1}_{3\times 3}
      &	0	&	0	&	0\\
      0	&\frac{G_{\s}}{ Z_{\sp\phi } } & -\frac{G_{\s\sp\D_1}}{  
        \sqrt{Z_{\sp\phi }Z_{\sp\Delta }} } & -\frac{G_{\s\sp\D_2}}{  
        \sqrt{Z_{\sp\phi }Z_{\sp\Delta }} } \\
      0	&  -\frac{G_{\s\sp\D_1}}{  \sqrt{Z_{\sp\phi }Z_{\sp\Delta }} }
      & \frac{G_{\sp\D_1}}{ Z_{\sp\D } }  & \frac{G_{
          \sp\D_1\D_2}}{ Z_{\sp\D } }\\
      0 	& \frac{G_{\s\sp\D_2}}{  \sqrt{Z_{\sp\phi }
          Z_{\sp\Delta }} } &- \frac{G_{\sp\D_1\D_2}}{ Z_{\sp\D } }& 
      \frac{G_{\sp\D_2}}{ Z_{\sp\D } } 
    \end{array}
  \right)\, \nonumber 
\eea 
with
\begin{align}
  G_\p(\vec{p}^{\,2})&=K_\p^{-1}\,, \
  G_\s(\vec{p}^{\,2})=\left(K_{\sp\D_1} K_{\sp\D_2}+16\m^2 p_0^2
  \right)/J \,, 
\end{align}
for the mesons and
\begin{align}
  G_{\sp\D_1}(\vec{p}^{\,2})&=K_{\sp\D_2}K_{\s}/J \,,\nonumber\\
  G_{\sp\D_2}(\vec{p}^{\,2})&=\left( K_{\sp\D_1}K_{\s}-4\rho_{\sp\phi
    } \rho_{\sp\Delta }V_{\sp\f\D}^{2} \right) /J\,, &\qquad&&
\end{align}
for the baryons. The indices $\phi$ and $\D$ of $V$ denote derivatives
with respect to $\rho_{\sp\phi }$ and $\rho_{\sp\Delta}$. Further, we
have the mixed contributions
\begin{align}
  G_{\s\sp\D_1}(\vec{p}^{\,2})&=2V_{\sp\f\D}\sqrt{\r_{\sp\f} 
\rho _{\sp\Delta } }K_{\sp\D_2} /J\,,\nonumber\\
  G_{\s\sp\D_2}(\vec{p}^{\,2})&=8V_{\sp\f\D}\sqrt{\r_{\sp\f} 
\rho _{\sp\Delta } }\m p_0 /J\,,\nonumber\\
  G_{\sp\D_1\D_2}(\vec{p}^{\,2})&=4\m p_0 K_\s/J\,,
\end{align}
where 
\bea J(\vec{p}^{\,2})=K_{\s}\left(K_{\sp\D_1} K_{\sp\D_2}+16\m^2
  p_0^2 \right)- 4\rho _{\sp\phi } \rho _{\sp\Delta
}V_{\sp\f\D}^{2}K_{\sp\D_2} \,,
\eea 
with 
\begin{align}
  K_\p(\vec{p}^{\,2})&= \vec{p}^{\,2}+{p}^2_0+
  V_{\sp\f}\,,\nonumber\\
  K_\s(\vec{p}^{\,2})&= \vec{p}^2+{p}^2_0+
  V_{\sp\f}+2 \r_{\sp\f} V_{\sp\f\f}\,,\nonumber\\
  K_{\sp\D_1}{(\vec{p}^{\,2})}&= \vec{p}^{\,2}+ {p}^2_0+V_{\sp\D}+2
  \r_{\sp\D} V_{\sp\D\D}-4\m^2\,,
  \nonumber\\
  K_{\sp\D_2}{(\vec{p}^{\,2})}&=
  \vec{p}^{\,2}+{p}^2_0+V_{\sp\D}-4\m^2\,,
\end{align}
is the determinant of the $\s\D$-part of $G_\vf$ without the $Z$'s. We
denote only space-like momentum dependencies since they will be
crucial for derivations of anomalous dimensions. For the derivation of
flow equations we will need derivatives of each entries of the
propagators with respect to the $K$'s. It can be checked that 
\bea \frac{\pa G_{\vf, ij} }{\pa K_{\vf_l}}=-G_{\vf, il}G_{\vf,
  lj}\,. \label{eq:KderivGvf} \eea
Note that equal indices do not imply a sum in \eq{eq:KderivGvf}.
Using this relations, one can express momentum derivatives
conveniently in terms of the propagators
\bea \frac{\pa G_{\vf, ij} }{\pa \vec{p}^{2}}=\sum_{l}\frac{\pa
  G_{\vf, ij} }{\pa K_{\vf_l}}=-\sum_{l}G_{\vf, il}G_{\vf,
  lj}\,. \label{eq:momderGvf} \eea
The boson propagator in the complex representation is
\bea
G_{\bar{\varphi}} = \left(
    \begin{array}{cccc}
      \frac{ G_\p}{Z_{\scriptscriptstyle \phi } }\,\mathds{1}_{3\times 3}
      &	0	&	0	&	0\\
      0	&\frac{G_{\s}}{ Z_{\sp\phi } } & \frac{G_{\s\sp\D}^-}{  
        \sqrt{Z_{\sp\phi }Z_{\sp\Delta }} } & \frac{G_{\s\sp\D}^+}{  
        \sqrt{Z_{\sp\phi }Z_{\sp\Delta }} } \\
      0	&  \frac{G_{\s\sp\D}^+}{  \sqrt{Z_{\sp\phi }Z_{\sp
            \Delta }} }  & \frac{G_{\sp|\D|}}{ Z_{\sp\D } } 
      & \frac{G_{\sp\D}^+}{ Z_{\sp\D } }\\
      0 	& \frac{G_{\s\sp\D}^-}{  \sqrt{Z_{\sp\phi }
          Z_{\sp\Delta }} } & \frac{G_{\sp\D}^-}{ Z_{\sp\D } }
      & \frac{G_{\sp|\D|}}{ Z_{\sp\D } } 
    \end{array}
  \right)\, \eea
with
\begin{align}
  K_{\sp\D }^\pm(\vec{p}^0) &=\vec{p}^{\,2}+\left( {p}_0\pm 2i\m
  \right) ^2+V_{\sp\D}\,,
  \nonumber\\
  G_{\sp\D}^\pm(\vec{p}^{\,2}) &= \left[ (K_{\sp\D}^\mp+ \r_{\sp\D}
    V_{\sp\D\D}) K_{\s}-2\rho _{\sp\phi }
    \rho _{\sp\Delta }V_{\sp\f\D} ^{2}\right] /J\,,\nonumber\\
  G_{\sp|\D|}(\vec{p}^{\,2})&=\left( 2 \rho _{\sp\phi }
    V_{\sp\f\D}^{2} -V_{\sp\D\D}K_{\s} \right)
  \rho _{\sp\Delta }/J\,,\nonumber\\
  G_{\s\sp\D}^\pm (\vec{p}^{\,2})&=-\sqrt{2 \rho _{\sp\f }\rho
    _{\sp\D}} \,V_{\sp\f\D} K_{\sp\D}^\mp/J\,. 
\end{align}
Note that the propagator is hermitian, i.e. $G_{\bar\vf,
  ij}^*=G_{\bar\vf, ji}$ and furthermore obeys $G_{\bar\vf,
  ij}(-p_0)=G_{\bar\vf, ji}(p_0)$.  Since the transformation between
representations affects only the diquark space, $G_\s$ does not
change, neither does the determinant. The three boson vertex is given
by
\begin{align}
  \G_{\vf_k\vf_j\vf_i} =& \quad Z_j\d_{ji}\left(Z_{\sp\f}\s V_{\sp \f
      j} \d_{k4} +
    Z_{\sp\D}\D V_{\sp j \D}  \d_{k5}\right) \nonumber\\
  & +Z_k\d_{kj}\left( Z_{\sp\f}\s V_{\sp \f k} \d_{i4} +
    Z_{\sp\D} V_{\sp k \D} \D\d_{i5}\right) \nonumber\\
  &+Z_i\d_{ik}\left(Z_{\sp\f}\s V_{\sp \f i} \d_{j4} +
    Z_{\sp\D}\D V_{\sp i \D} \d_{j5}\right) \nonumber\\
  & +\left( Z_{\sp\f}\s\d_{k4}+Z_{\sp\D}\D\d_{k5}\right) \left(
    Z_{\sp\f}\s\d_{i4}+ Z_{\sp\D}\D\d_{i5}\right)
  \nonumber\\
  &\qquad \times \left( Z_{\sp\f}\s\d_{j4}+
    Z_{\sp\D}\D\d_{j5}\right)V_{ijk} \,,
\end{align}
where, apart from the Kronecker delta, we apply the convention that
the index is set to $\f$ if it corresponds to a mesonic field, and to
$\D$ if it corresponds to diquark field in $\vf$. Accordingly
$V_{ijk}$, where the indices denote $\r$-derivatives, can either have
three derivatives with respect to one species or two with respect to
one and one with respect to the other.
\subsection{Fermions}\label{app:NGP}
\vspace{-3.mm}

It is convenient to apply the Nambu-Gorkov formalism, where the quark
fields are represented by the bi-spinors in color space 
\bea
\Psi=\left({{\begin{array}{c} q_{r} \\
        \t_{2}C\bar{q}^T_{g}\end{array}}}\right)\,,\quad
\bar{\Psi}=\left({{\begin{array}{c}\bar{ q}_{r} \\
        q^T_{g}C\t_{2}\end{array}}}\right), 
\eea 
where the indices $r$ and $g$ denote the color components of the quark
spinor. In the convention we apply here, the Nambu-Gorkov space is
equivalent to color space.  Thus we can rewrite the quark part of the
effective action as 
$ \G_{q} =\int_x \bar{\Psi}(x) S(x,i\pa)
\Psi(x) \label{eq:NGrepOfG} 
$ 
with
\begin{widetext}
\begin{eqnarray}
  S(x,i\pa)=
  Z_q
  \left({{\begin{array}{cc} 
          i\slashed{\pa}+i\g_{0}\m+i\sqrt{Z_{\sp \f}}h_{\sp\f} 
          \left[\s(x)+i\g_5\vec{\t}\cdot\vec{\p}(x) \right]& -
          \sqrt{2Z_{\sp\D}}h_{\sp\D}\Delta(x)\gamma_5\\
          - \sqrt{2Z_{\sp\D}} h_{\sp\D}\Delta^{*}(x)\gamma_5 
          &i\slashed{\pa} -i\g_{0}\m+i\sqrt{Z_{\sp\f}} h_{\sp\f}
          \left[\s(x)-i\g_5\vec{\t}\cdot\vec{\p} (x)\right]
\end{array}}}\right)\,.\label{eq:NambuGorkov} 
\end{eqnarray}
\end{widetext}
Now we write the inverse propagator at the expansion point as
\begin{eqnarray}
&&G_{\J\bar{\J}}^{-1}(p)=-\big[G_{\bar{\J}\J}^{-1}(-p)\big]^T\nonumber\\
&=&Z_q\left(
          \begin{array}{cc}
            \slashed{\vec{p}}+\g_0 (p_{0} +i\m)+i \hat{\s} &
            - \hat{ \D } \g_5 \\
            - \hat{ \D} \g_5  &   {\slashed {\vec{p}}}+\g_0 (p_{0} 
            -i\m)+i\hat{\s}
              \end{array}
             \right)\,.  \nonumber 
\end{eqnarray}
Note that if there is no tensor structure regarding a particular space
(color/flavor/spinor space) an identity matrix with respect to that
specific space is implied.  Furthermore we choose the diquark
condensate to be on the real axis. Now we invert above equation to
find the Nambu-Gorkov Propagator 
\begin{eqnarray} { G_{\Psi\bar{\Psi}}}
  = \frac{1}{Z_q}\left(
          \begin{array}{cc}
            G^+  &  \Delta^{-}\\
            \Delta^{+} & G^-
               \end{array} \right) 
\, , 
\end{eqnarray}
where 
\begin{eqnarray}\nonumber 
  G^{\pm}
  &=&\left(\slashed{\vec{p}} -i \hat{\s} \right)A_{\pm}
  +\g_0 B_{\pm} \,,
  \\[2ex]
  \Delta^{\pm}
  &=& \g_5 \left[ \hat{\D}\,A\pm i F (\slashed{\vec{p}} 
    +i\hat{\s})\g_0\right]\label{eq:Dpm}\,.
\end{eqnarray}
 With 
$
\nu_\pm=p_0\pm i \m
$
we define the functions as
\begin{align}
  A(\vec{p} ^{\,2})&= \left(\vec{p}^{\,2}+\nu_+\nu_-+\hat{\s}^2
    +\hat{\D}^2\right)D(\vec{p} ^{\,2})\,,
  \nonumber\\
  A_\pm(\vec{p}^{\,2})&= \left(\vec{p}
    ^{\,2}+\nu_\mp^2+\hat{\s}^2+\hat{\D}^2 \right)D(\vec{p} ^{\,2})\,,
  \nonumber\\
  B_\pm(\vec{p} ^{\,2})&=\left(\nu_\pm\left(\vec{p}
      ^{\,2}+\nu_\mp^{2}+\hat{\s}^2\right)+\nu_\mp
    \hat{\D}^2\right)D(\vec{p} ^{\,2})\,, 
\end{align}
 and
\begin{align}
F(\vec{p} ^{\,2})&=  2\hat \D\m D(\vec{p} ^{\,2})\, ,&
D(\vec{p} ^{\,2})&=\frac{1}{ a_+ a_-+ f^{2 }} \,,
\end{align}
where the small letter functions are the numerators of the capital
letter functions. These functions have the following properties, for
the sake of demonstration, we explicitly show here the
$p_0$-dependence and omit the $\vec{p}^{\,2}$-dependence
\begin{align}
  A(-p_0)&=A(p_0),& A_\pm(-p_0)&=A_\mp(p_0),\nonumber\\
  B_\pm(-p_0)&=-B_\mp(p_0),&
  F(-p_0)&=F(p_0)\,, 
\end{align}
as well as
\begin{align}
  A^{*}&=A,& A_\pm^{*}&=A_\mp,& B_\pm^{*}&= B_\mp,&
  F^{*}&=F\,.
\end{align}
For the flow equations, the following derivatives will be required
 \begin{align}
A'_{\pm}&=F^{2}-A_{\pm}^{2}\,, &
A'&=D-A(A_{+}+A_{-})\,,\nonumber\\
 B'_{\pm}&=\nu_{\pm}D-B_{\pm}(A_{+}+A_{-})\,,& 
F'&=-F(A_{+}+A_{-})\,,
 \end{align}
 where primes denote a derivation with respect to $\vec{p}^{\,2}$. Let
 us turn to the vertices. The quark-meson ones are given by
\begin{align}
  \G_{\bar{\J}\s \J}&=\sqrt{Z_\f}Z_q ih_{\sp\f} \;\mathds{1}\,,
  \nonumber\\
  \G_{\bar{\J}\p_i \J}&=\sqrt{Z_\f}Z_qh_{\sp\f}
  \left({{\begin{array}{cc}
          -\gamma_5 \tau_i& 0\\
          0 & \gamma_5\tau_i
\end{array}}}\right)\,.
\end{align}
The quark-diquark vertices read
\begin{align}
  \G_{\bar{\J}{\sp\D} \J}&=\sqrt{2Z_{\sp\D}}Z_q h_{\sp\D}
  \left({{\begin{array}{cc}
          0& -\gamma_5\\
          0 & 0
        \end{array}}}\right)\,,\nonumber\\
  \G_{\bar{\J}{\sp\D}^{*} \J}&=\sqrt{2Z_{\sp\D}}Z_q h_{\sp\D}
  \left({{\begin{array}{cc} 
          0&0\\
          -\gamma_{5} & 0
\end{array}}}\right)
\,.
\end{align}
In the real representation they are 
\begin{align}
  \G_{\bar{\J}{\sp\D}_{1} \J}&=\sqrt{Z_{\sp\D}}Z_q
  h_{\sp\D}\left({{\begin{array}{cc}
          0& -\gamma_5\\
          -\gamma_5 & 0
        \end{array}}}\right)x
  \,,\nonumber\\
  \G_{\bar{\J}{\sp\D}_{2} \J}&=\sqrt{Z_{\sp\D}}Z_q 
  h_{\sp\D}\left({{\begin{array}{cc} 
          0& -i \gamma_{5}\\
          i \gamma_{5} & 0
\end{array}}}\right)\,.
\end{align}
The unit matrix
$\mathds{1}=\mathds{1}_{f}\times\mathds{1}_{c}\times\mathds{1}_{D}$ is
composed of a tensor product of unit matrices in color, flavor and
Dirac space. Again, if there is no tensor structure regarding a
particular space, an identity matrix is implied.

For completeness, we show the projection operator for the diquark
Yukawa coupling we use in Sec.~\ref{ssec:Yukawa},
\begin{align}\label{eq:hatp}
 \hat{P}=
\left(
\begin{array}{ccc}
  0 &  -\g_{5}   \\
  -\g_{5}  &    0
\end{array}
\right)\,.
\end{align}
It has the tensor structure of the real part of the quark-diquark
vertex, since the diquark background field is real-valued.

\section{Flow equations}\label{app:Fleq}
\vspace{-3.mm}

We employ the 3d flat regulator functions \cite{Litim:2001up} for
bosons and fermions
 \begin{align}
r_{\sp B}(x)&=\left(\frac{1}{x}-1 \right) \Theta (1-x) \,,
\nonumber\\[2ex]
r_{\sp F}(x)&=\left(\frac{1}{\sqrt{x}}-1 \right) \Theta (1-x)\,. 
\label{eq:regs}\end{align}
The formal scale derivative acts only on the regulators
 \begin{align}\nonumber 
   \tilde{\pa_t} r_{\sp B} &=\left[ \frac{2}{x} - \eta_{\sp
       B}\left(\frac{1}{x} - 1\right)
   \right]\Theta(1-x)\,,\\[2ex]
   \tilde{\pa_t} r_{\sp F} &=\left[ \frac{1}{\sqrt{x}} - \eta_{\sp
       F}\left(\frac{1}{\sqrt{x}} - 1\right)\right]\Theta(1-x)\,,
\label{eq:dtr}
\end{align}
where we considered that the regulator $R_{k}$ contains the wave
function renormalisations. The flow equations will be displayed in
terms of the following integrals, where we apply the shorthand
notation $\int_{q}=T\sum_{n}\int \frac{d^{3}q}{(2\p)^{3}}$. The sum is
over the Matsubara frequencies. The bosonic anomalous dimension
contains two types of integrals. The first one yields the same results
for bosonic and fermionic integrands
\begin{widetext}
\begin{align}
  I^{(1)}_{\h_{\sp B}}[G,H]=\tilde{\pa}_t
  \frac{\pa}{\pa\vec{p}^2}\int_q G(\vec{q}^2_r)
  H((\vec{q}+\vec{p})^2_r)\Big\lvert_{p=0}=\frac{k^5 T}{3\p^{2}}
  \sum_n {G}'(k^2){H}'(k^2)\, .
\end{align}
Derivatives of fermionic functions are given in
App. \ref{app:NGP}. Bosonic derivatives can be performed with
\eq{eq:momderGvf}. The second integral only occurs for
fermionic integrands
\begin{align}
  I^{(2)}_{\h_{\sp B}}[G,H,m^{2}]&= \tilde{\pa}_t
  \frac{\pa}{\pa\vec{p}^2}\int_q G(\vec{q}^2_r)
  H((\vec{q}+\vec{p})^2_r) \left[\vec{q}_r \cdot (\vec{q}+\vec{p})_r+
    m^{2}\right]\Big\lvert_{p=0}\nonumber\\
  &=\frac{k^3 T}{3\p^{2}} \sum_n \left[ k^2(k^2+m^2) {G}'
    (k^2){H}'(k^2)-\frac{ {G}(k^2) {H}(k^2)}{4}-\left( 1-\h \right)
    {G}(k^2)\left(\frac{ {H}(k^2)}{2}+ k^2 {H}'(k^2)\right) \right]\,.
\end{align}
This is only valid for one single fermionic species. Fermionic
self-energy diagrams contain mixed integrals
\begin{align}
  I_{\h_{\sp F}}[G_{\sp F},H_{\sp B}]=-\tilde{\pa}_t \frac{\pa}{\pa
    \vec{p}}\cdot\int_q \vec{q}(1+r_{\sp F}(\vec{q}^{2}))G_{\sp
    F}(\vec{q}^2_{r}) H_{\sp
    B}((\vec{q}-\vec{p})^2_{r})\Big\lvert_{p=0}=-\frac{k^{5}T}{\p^{2}}\sum
  _{n} G_{\sp F}(k^2) \sum_{i} \left(1- \frac{\h_{\sp
        \vf_i}}{4}\right)\frac{\pa H_{\sp B}(k^2)}{\pa
    K_{\vf_{i}}(k^2)}\,.
\end{align}
The second sum is over all bosonic species and $K_{\vf_{i}}$
represents the functions given in App. \ref{app:BOSONpropagator} in
particular \eq{eq:KderivGvf} can be used.
And finally 
\begin{align}
  I_{h}[G_{\sp F},H_{\sp B}]=-\tilde{\pa}_t \int_q G_{\sp
    F}(\vec{q}^2_{r}) H_{\sp
    B}(\vec{q}^2_{r})=-\frac{k^{5}T}{3\p^{2}}\sum _{n}\bigg[ G_{\sp F}
  (k^2) \sum_{i} \left(1- \frac{\h_{\sp \vf_i}}{4}\right)\frac{\pa
    H_{\sp B}(k^2)}{\pa K_{\vf_{i}}(k^2)}+\left(1-\frac{\h_{\sp F}}{4}
  \right) H_{\sp B}(k^2)G'_{\sp F}(k^2)\bigg]\,.  
\end{align}
Now can we write down all flow equations in terms of the above
integrals. The anomalous dimensions are
\begin{align}
  \h_{\sp\f}&= 2I^{(1)}_{\h_{\sp B}}\left[{G_\p},\r_{\sp\f}
    V_{\sp\f\sp\f}^2 {G}_{\s} +\r_{\sp\D}V_{\sp\f\sp\D}^2
    {G}_{\sp\D_1} -2\sqrt{ \r_{\sp\f} \r_{\sp\D}}\,
    V_{\sp\f\sp\f}V_{\sp\f\sp\D}G_{\s\sp\D_1} \right]
  \nonumber\\&\quad
  +4 N_c N_f h_{{\f}}^2 \left\lbrace I^{(2)}_{\h_{\sp
        B}}\left[A_+,A_+,\hat{\s}^2\right] + I^{(1)}_{\h_{\sp
        B}}\left[B_+,B_+\right]+ \hat{\D}^2 I^{(1)}_{\h_{\sp
        B}}\left[A,A\right] - I^{(2)}_{\h_{\sp
        B}}\left[F,F,\hat{\s}^2\right]\right\rbrace\,,
  \nonumber\\
  \h_{\sp\D}&= 2 I^{(1)}_{\h_{\sp
      B}}\left[G_{\sp\D_2},\r_{\sp\D}V_{\sp\D\D}^2 {G}_{\sp\D_1}
    +\r_{\sp\f} V_{\sp\f\D}^2 {G}_{\s} -2\sqrt{ \r_{\sp\f}
      \r_{\sp\D}}\, V_{\sp\D\D}V_{\sp\f\sp\D}G_{\s\sp\D_1}\right]
  +2\r_{\sp\f}V_{\sp\f\D}^2 I^{(1)}_{\h_{\sp
      B}}\left[G_{\s\sp\D_2},G_{\s\sp\D_2}\right] \nonumber\\&\quad
  +2\r_{\sp\D}V_{\sp\D\D}^2 I^{(1)}_{\h_{\sp
      B}}\left[G_{\sp\D_1\D_2},G_{\sp\D_1\D_2}\right]
-4\sqrt{\r_{\sp\f}\r_{\sp\D}}\,V_{\sp\D\D}V_{\sp\f\D}  
I^{(1)}_{\h_{\sp B}}\left[G_{\s\sp\D_2},G_{\sp\D_1\D_2}\right]
\nonumber\\&\quad
+4 N_c N_f h_{\D}^2 \left\lbrace I^{(2)}_{\h_{\sp
      B}}\left[A_-,A_+,\hat{\s}^2\right] + I^{(1)}_{\h_{\sp
      B}}\left[B_-,B_+\right]+ \hat{\D}^2 I^{(1)}_{\h_{\sp
      B}}\left[A,A\right] + I^{(2)}_{\h_{\sp
      B}}\left[F,F,\hat{\s}^2\right]\right\rbrace\,,
\nonumber\\
\h_{q}&=\frac{1}{3}h_{\sp\f}^{2}\;\Re\,I_{\h_{\sp F}}\left[A_{+},
  3G_{\p}+G_{\s}\right]+\frac{2}{3}h_{\sp\D}^{2}\;\Re\,I_{\h_{\sp
    F}}\left[A_{+}, G_{\sp\D}^{+}\right]\,.
\end{align}
We note that, at finite temperature, the components of the wave
function renormalisations parallel, $Z^{\parallel}$, and
perpendicular, $Z$, to the heat bath are different a priori. However,
at scales above the temperature scale, $T/k\!<\!1$, the system is
insensitive to thermal effects. On the other hand, for $T/k\gg1$ the
finite temperature RG flow is only driven by the lowest Matsubara
mode. While it is zero for bosons and $Z^{\parallel}$ drops out, it is
proportional to $T$ for fermions which therefore decouple in this
regime. Hence, we used the approximation $Z^{\parallel}\equiv Z$.
Finally, the flow equations of the Yukawa couplings are
\begin{align}\nonumber 
 {\pa}_t h_{\sp\f} 
  &= \left(\h_{q}+\frac{\h_{\sp\f}}{2}\right)h_{\sp\f}+ h_{\sp\f}
  \Re\left\{ h_{\sp\f} ^{2}I_{h}\left[A_{+}, 3G_{\p}(q)-G_{\s}(q)
    \right]+2h_{\sp\D}^{2}I_{h}\left[A_{+},
      G_{\sp\D}^{+}\right]\right\}\,,
  \\
  \quad{\pa}_t h_{\sp\D}
  &= \left(\h_{q}+\frac{\h_{\sp\D}}{2}\right)h_{\sp\D}+h_{\sp\D}
  \Re\left\{ h_{\sp\f} ^{2}I_{h}\left[A_{}, 3G_{\p}(q)+G_{\s}(q)
    \right] -2h_{\sp\D}^{2}I_{h}\left[A_{}, G_{\sp|\D|}^{}\right] - 2
    \frac{h_{\sp\f}^{2}\s}{\D}
    I_{h}\left[A_{+},G_{\s\sp\D}^{+}\right]\right\}\,.
\end{align}
\end{widetext}

%%%%%%%%%%%%%%%%%%%%%%%%%%%%%%%%%%

\section{One-dimensional Taylor expansion}\label{app:1dt}

\vspace{-3.mm}

%%%%
%%%%%%%%%%%%%%
\begin{figure}[t]
  \includegraphics[width=.9\columnwidth]{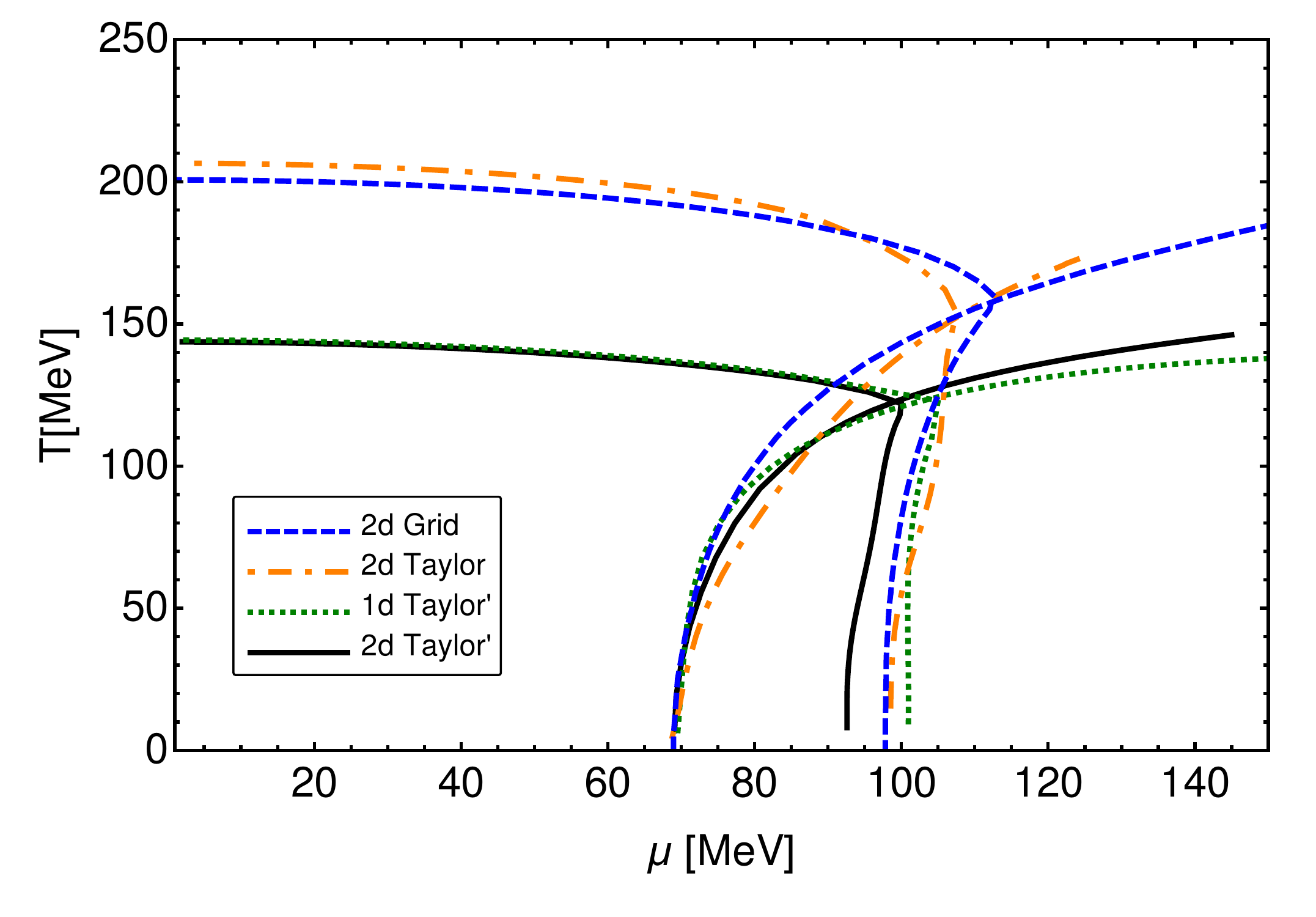}
  \caption{A comparison between various phase diagrams obtained from
    different truncations with the FRG. We want to emphasise that
    there is only a small difference between the one-dimensional
    Taylor expansion (dotted green line) and the two-dimensional
    Taylor expansion (solid black line).}
   \label{fig:allfrgpds}
\end{figure}
%%%%%%%%%%%%%%
%%%%

We mentioned in Sec.~\ref{ssec:flowofparam} that with a
one-dimensional expansion of the effective potential in terms of the
$SO(6)$-invariant $\r=\r_{\sp\f}+\r_{\sp\D}$ the violation of the
Silver Blaze property for an expansion scheme in terms of momentum
independent meson-diquark correlation functions can be avoided. The
reason is that the $SO(6)$ flavor symmetry relates all
$2n$-meson--$2m$-diquark interactions $\lambda_{n,m}$ to purely
mesonic interactions, $\lambda_{n,m} = \lambda_{n+m,0}$. Hence, all
$n$-point functions that involve baryons can be defined purely in
terms of mesons. This bypasses the necessity to consider frequency
dependent couplings.

This construction assumes a larger flavor symmetry than the system has
at finite density. It is nonetheless instructive to consider the
one-dimensional expansion since technically less demanding than the
two-dimensional expansion. Furthermore, it allowed us to explicit
confirm our statements concerning the Silver Blaze property. Thus, we
will provide the parameterisations as well as a set of initial
conditions for the one-dimensional Taylor expansion here.

In the normal phase with $\r_{\sp\D}\!=\!0$ we parametrise the
effective potential as
\begin{align}
  V_{nor}(\r) =m^2 \r +\sum_{n=2}^{N} \frac{\l_n}{n!}(\r-
  \k)^n\,, \label{eq:1dtaylor}
\end{align}
where $\k=\k_{\sp\f}+ \k_{\sp\D}$ and we dropped irrelevant
constant. With this Ansatz the derivative terms in \eq{eq:min} reduce
to $V'_{nor}(\k)=m^2$ and the solution is given by
\bea \k_{\sp \f}=\frac{1}{2}\left(\frac{c}{m^2}\right)^2 , \;\;\;
\k_{\sp\D}=0\, .\label{eq:condnor} \eea
Note that this solution is only valid away from the chiral limit
$c\!=\!0$.  For the effective potential to have a minimum, the
determinant of the Hessian matrix needs to be larger than zero, 
\bea \det\left(\frac{\pa^2 U}{\pa\varphi_i
    \pa\varphi_j}\right)\bigg\lvert_{\r=\k}=\left(m^2
\right)^{N_{\sp\f}}\left( m^2-4\mu
  ^2\right)^{N_{\sp\D}}>0\,,\label{eq:hesdet} \eea
where $\vf_{i}$ represents the real-valued components of the boson
field (see App. \ref{app:BOSONpropagator}), and $N_{\sp\f},N_{\sp\D}$
are the number of meson and diquark fields respectively. As long as $
m^2>4\m^2$ we are in the normal regime. At $m^2=4\m^2$ the diquark
curvature mass vanishes. This indicates a second order phase
transition to the BEC phase and spontaneous breaking of $U(1)_B$
symmetry. For $m^2<4\m^2$ we use the parametrisation discussed below.

Since $m$ is the mass of the pions, diquark condensation sets in when
the quark chemical potential reaches half of the pion mass. The latter
is, according to the Silver Blaze property, independent of the
chemical potential at $T=0$. Hence, if the Silver Blaze property is
not violated, $\m_{c}=m_{\p}/2$ holds within the 1d-Taylor expansion
Ansatz. We will elaborate on this in Sec.~\ref{sec:SilvBl}.

For the BEC phase we reparameterise the potential by renaming $m^2=
\l_2(\k-v)$, thus we exchange $m$ in favor of $v$ and write (by adding
an irrelevant constant)
\be V_{bec}(\r)=\frac{\l_2}{2}(\r-v)^2+\sum_{n=3}^{N}
\frac{\l_n}{n!}(\r- \k)^n\, .\label{eq:1dtaylorBEC} \ee
In this parametrisation we have \mbox{$V'_{bec}(\k)= \l_2(\k-v)$}, and
\eq{eq:min} now yields
\begin{align}
  \k_{\sp \f}=\frac{1}{2}\left(\frac{c}{ 4\m^2}\right)^2 \,,\quad
  \k_{\sp\D}=v+\frac{ 4 \m^2}{ \l_2}-\frac{1}{2}\left(\frac{ c}{4
      \m^2}\right)^2\, .\label{eq:vevBEC}
\end{align}
We see that the chiral order parameter $\s_{0}=\sqrt{2 \k_{\sp \f}}$
is proportional to $\mu^{-2}$. Furthermore, with rising $\mu$ the
ground state rotates from the mesonic direction to the diquark
direction.

%%
%%%%%%%%%
 \begin{figure}[t]
\includegraphics[width=.9\columnwidth]{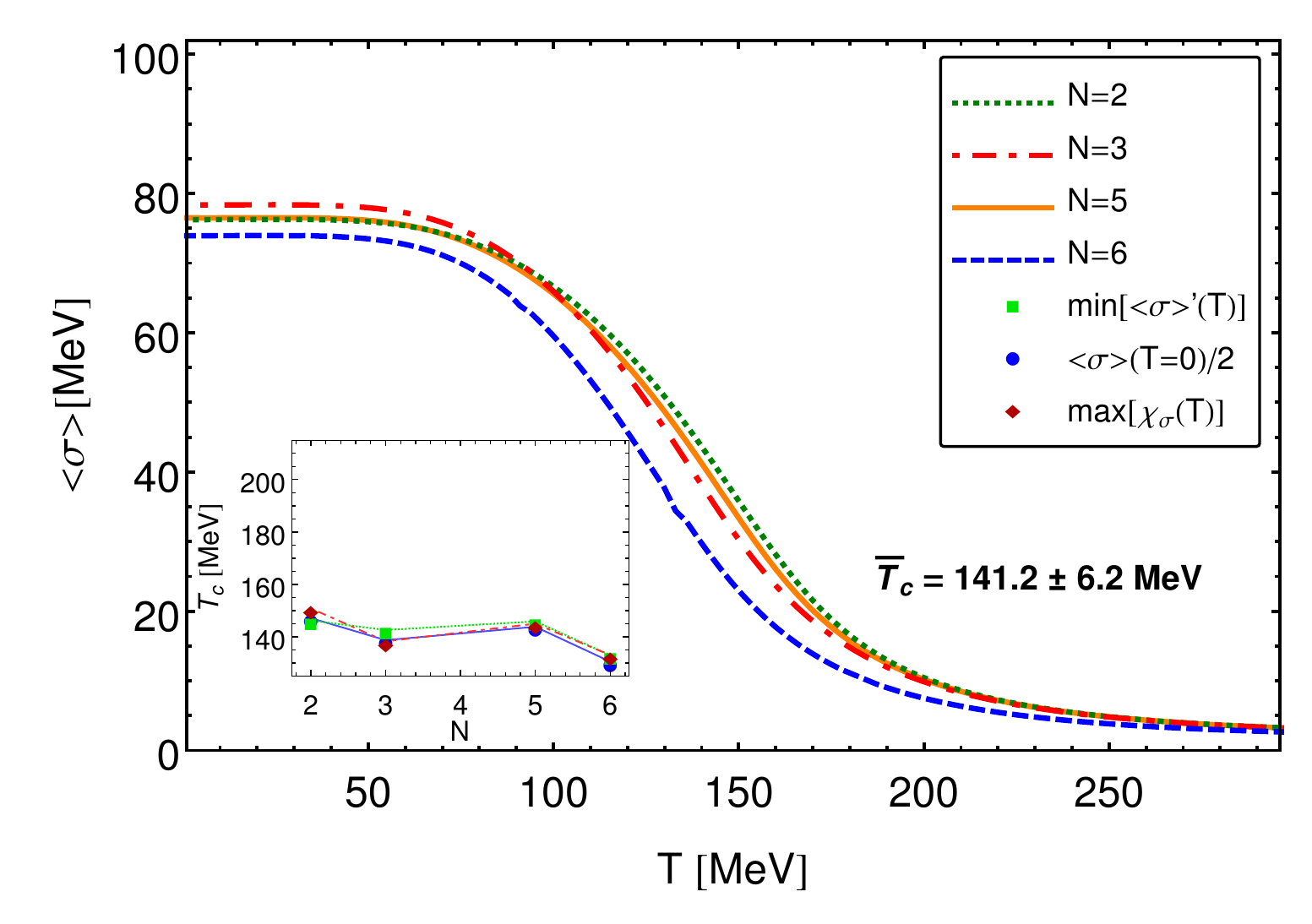}
\caption{Chiral condensate over temperature as well as the transition
  temperatures at $\m=0$ for various orders of the expansion of the
  effective potential. }
\label{fig:Conv}
    \end{figure}
%%%%%%%%%
%%

Furthermore, we set $h_{\sp\D}=h_{\sp\f}$, and following our
discussion above, the flows of all Taylor coefficients are given by
the mesonic ones. However, the wave function renormalisations of
mesons and diquarks are independent. For the UV cutoff $\Lambda =
900\,\text{MeV}$ we used as initial parameters $\langle
\sigma\rangle_\Lambda \!=\! 2.28\, \text{MeV}$, $m_{\phi,\Lambda}
\!=\! 1090\, \text{MeV}$, $\lambda_{2,\Lambda} \!=\! 89$ and
$h_{\phi,\Lambda} \!=\! 6.3$. These initial conditions give a
constituent quark mass of $360\,\text{MeV}$,
$m_\pi\!=\!139\,\text{MeV}$ as well as $f_\pi\!=\!76\,\text{MeV}$ in
the IR. Since Silver Blaze is fulfilled by construction in this case,
the latter exactly determines the onset chemical potential,
i.e. $m_\pi\!=\!2\mu_c$.

The resulting phase diagram is shown in
Fig.~\ref{fig:allfrgpds}. Remarkably, there is only a small difference
between the two-dimensional expansion (sold black line) and the
one-dimensional expansion (dotted green line). Hence, the dynamics of
QC${}_2$D are mainly driven by the mesons, also at high density. This
may be the most striking difference between QC${}_2$D and QCD but has
yet to be explored.

Finally, we discuss the convergence of the 1d Taylor expansion.  In
Fig.~\ref{fig:Conv} we show the chiral condensate as a function of the
temperature for different orders $N$ of the Taylor expansion. All
curves are computed with the same UV initial conditions as specified
above. We see that the expansion does not converge for the orders we
have computed. The crossover temperature has been computed with three
different definitions as shown in the legend of
Fig.~\ref{fig:Conv}. As a function of the order $N$ the critical
temperature does not converge as well. Therefore the average
$\bar{T}_c$ computed from the arithmetic mean of all three definitions
of $t_c$ and all orders, together with its margin of error, should
give a feeling for the error of our phase boundaries.

    In Ref.~\cite{Pawlowski:2014zaa} a Taylor expansion with a
    fixed expansion point, rather than a co-moving minimum, was
    employed for a quark-meson model. There, a rapid convergence of
    the expansion was observed for all temperature regions. As it was
    discussed there, the back-coupling of higher order couplings into
    the flow equations at lower orders in the co-moving expansion is
    responsible for numerical instabilities and slow
    convergence. Hence, in light of these findings, the lack of
    convergence in the present work is no surprise.

\end{appendix}

\bibliography{qc_2d}

\end{document}